\documentclass[preprint, 12pt]{elsarticle}

\usepackage{amssymb}
\usepackage{xcolor}
\usepackage{amsmath}
\usepackage{physics}
\usepackage{subcaption}

\journal{Computers \& Fluids}

\begin{document}

\begin{frontmatter}

\title{Improved Delayed Detached Eddy Simulation with Reynolds-Stress Background Modeling}

\author[1]{Marius Herr}
\ead{m.herr@tu-braunschweig.de}
\author[1]{Rolf Radespiel}
\ead{r.radespiel@tu-braunschweig.de}
\author[2]{Axel Probst}
\ead{axel.probst@dlr.de}

\affiliation[1]{organization={Technical University Braunschweig, Institute of Fluid Mechanics},%
            addressline={Hermann-Blenk-Str. 37}, 
            city={Braunschweig},
            postcode={38108}, 
            state={Lower Saxony},
            country={Germany}}

\affiliation[2]{organization={German Aerospace Center, Institute for Aerodynamics and Flow Technology},%
            addressline={Bunsenstr. 10}, 
            city={Göttingen},
            postcode={37073}, 
            state={Lower Saxony},
            country={Germany}}

\begin{abstract}
A novel variant of Improved Delayed Detached-Eddy Simulation based on a differential Reynolds-stress background model is presented.
The approach aims to combine the advantages of anisotropy-resolving Reynolds-stress closures in the modelled RANS regions with consistent LES and wall-modelled LES behaviour in the resolved flow regions.
In computations of decaying isotropic turbulence with a low-dissipative flow solver it is shown that a straightforward hybridised Reynolds-stress model provides insufficient turbulent dissipation as sub-grid closure in the LES regions and is therefore locally replaced by scalar viscosity modelling.
Simulations of periodic channel flows at different Reynolds numbers and grid resolutions are used to calibrate and validate the wall-modelled LES branch of the new model.
A final application in embedded wall-modelled LES of a flat-plate boundary layer is widely consistent with results using the SST-RANS background model, but shows some deviations from the Coles-Fernholz skin-friction correlation.
In this regard, initial sensitivity studies indicate possible adverse effects due to the synthetic-turbulence approach used in these simulations.

\end{abstract}


\end{frontmatter}

\section{Introduction}
Hybrid RANS-LES methods (HRLM) aim to combine the efficiency of statistical turbulence modelling (RANS) and the accuracy of scale-resolving simulations (LES) in order to improve the prediction of complex high-Reynolds number flow at feasible computational effort.
Among the numerous approaches to couple RANS and LES \cite{Frohlich2008}, the Detached-Eddy Simulation (DES) \cite{spalart1997comments} and its more recent variants (DDES \cite{Spalart2006}, IDDES \cite{shur2008hybrid}) have become particularly popular, mainly due to their rather straightforward implementation in unstructured flow solvers and applicability to even complex geometries and flows, see e.g. \cite{Masini2020a}, \cite{Probst2022}.

As pointed out by many research groups over the last years, the success
of hybrid RANS-LES
depends on both modelling and numerical aspects, e.g. non-zonal \cite{spalart1997comments} versus zonal methods \cite{Deck2012}, the robustness of the so-called \emph{RANS shielding} \cite{Menter2018}, \cite{Deck2020}, wall-modelled LES capabilities \cite{Nikitin2000}, \cite{shur2008hybrid}, the so-called \emph{grey-area issue} at the RANS-LES interface \cite{Mockett2015}, \cite{Kok2017}, and the derivation of accurate and robust discretization schemes for local LES 
\cite{Probst2016}, \cite{Pont2017}.

Interestingly, the choice of the underlying RANS turbulence model has gained relatively little attention, as all of the above cited publications rely on well-established eddy-viscosity models from the 1990s, i.e. the Spalart-Allmaras \cite{Spalart1992} or Menter's SST \cite{Menter1994} model.
Common arguments for this are, that these models are still considered as the state of the art, or that the overall accuracy of hybrid methods should depend only marginally on the RANS regions, or that more sophisticated RANS models may introduce numerical instabilities.
However, recent advances in the robust application of anisotropy-resolving differential Reynolds-stress models (RSM) \cite{Eisfeld2016} \cite{Jakirlic2016} \cite{SPORSCHILL2022108955} motivate us to revisit the potential benefits and specific requirements of combine such models with advanced hybrid RANS-LES methods like the Improved Delayed Detached-Eddy Simulation (IDDES).

In this regard, it should be noted that typical applications of DES-type HRLM have shifted away from flows with clear distinction of attached (RANS) and massively-separated (LES) flow (e.g. stalling airfoils or landing gears with massive separations) towards more complex mixtures of e.g. mildly-separated regions, vortical flows that interact with (attached) boundary layers and local corner flows in wing-body or nacelle-pylon junctions.
All the latter phenomena are for example present on aircraft in high-lift configuration.
In such complex applications, the underlying RANS model will have to model at least (initial) parts of these phenomena, as the switch to LES may be delayed or ambiguous due to insufficient local grid resolution or hybrid-modelling issues (e.g. the mentioned \emph{grea area}).

Therefore, a differential RSM that naturally captures the effects of streamline curvature, rotation and secondary motion should offer clear benefits for such flows not only when applied in pure RANS computations, but also as part of hybrid RANS-LES.
Another advantage of RSM comes from providing more realistic anisotropic stresses as input for synthetic-turbulence generators (STG), which are used in the framework of embedded (wall-modelled) LES, e.g. \cite{francois2015forced}.
Finally, certain HRLM approaches deliberately increase the importance of the underlying turbulence model by placing large RANS zones around critical flow regions. 
Notably, the \emph{Automatic Zonal DES} (AZDES) \cite{Ehrle2020} relies on an RSM to cover the whole shock region on airfoils in buffet conditions, while the airfoil wake is resolved with LES.

Although several authors have applied RSM in the framework of HRLM \cite{Probst2011, Maduta2012, Ehrle2020, Zhuchkov2016a}, they mostly focused on rather basic couplings with Delayed DES or SAS (scale-adaptive simulation) which are not suitable for (embedded) wall-modelled LES.
An early attempt by Chaouat et al. \cite{chaouat2005new}, based on the partially integrated transport model (PITM), transferred the advantages of RSM to the sub-grid scale level, thus achieving a more accurate description of flow anisotropy than with eddy viscosity sub-grid models.
A further attempt \cite{Probst2015a} to use RSM with IDDES to simulate a periodic channel flow in combination with a suitable low-dissipative numerical scheme showed some potential, but the overall approach was not sufficiently validated for fundamental test cases, like decaying isotropic turbulence (DIT).
More recently, Wang et al. \cite{Wang2021} published an IDDES based on the SSG/LRR-$\omega$ RSM of Eisfeld et al. \cite{Eisfeld2016} (also used in this work), following a similar direct coupling strategy as originally proposed by Probst et al. ~\cite{Probst2011} for RSM-based DDES and later adopted by \cite{Probst2015a} and e.g. \cite{Ehrle2020}.
However, as will be shown below, such a sub-grid model that directly relies on the (reduced) anisotropic stresses from the differential RSM equations, will not provide the required dissipation on the small scales to correctly model the turbulent energy cascade.
While this can be compensated by more dissipative numerical schemes (as e.g. done in ~\cite{Probst2011}), the accuracy of IDDES in WMLES mode strongly relies on low-dissipation schemes, as e.g. pointed out by \cite{Probst2016}, \cite{Carlsson2023}.
In this regard, it should be noted that Wang et al. \cite{Wang2021} do not provide fundamental validation studies of their RSM-IDDES in WMLES mode, e.g. for the periodic channel flow.

With this background, this present paper proposes a consistent hybrid coupling of RSM with IDDES, which is validated for both free turbulence (LES mode) and wall-bounded turbulent flow (WMLES mode) in the framework of a low-dissipative low-dispersive flow solver.
To achieve this, it is shown that the straightforward hybridisation of the differential RSM needs to be abandoned in favour of a local switch to eddy-viscosity-based modelling in the LES regions. 
Note that the DDES branch of IDDES, which mainly relies on the \emph{RANS shielding} function $f_d$ is not focus of this paper.
Although based on a different RSM, we preliminary adopt the adjusted formulation of this function from the RSM-DDES proposed in \cite{Probst2011}.

The paper is organised as follows: 
In Sec.~\ref{sec:deriv_hrlm} the modelling concept and the equations of the novel combination of the SSG/LRR-RSM \cite{Eisfeld2016} with IDDES \cite{shur2008hybrid} are presented.
Sec.~\ref{sec:num_meth} describes the numerical methods applied in this work, including the flow solver DLR-TAU \cite{Schwamborn2006} with its low-dissipation low-dispersion (LD2) scheme \cite{Probst2016}, as well as the synthetic turbulence generation and injection method used for embedded WMLES.
The subsequent Sec.~\ref{sec:results} presents validation results for three canonical flows covering different modelling regimes, i.e. decaying isotropic turbulence for pure LES, the periodic channel flow at different Reynolds numbers for general WMLES and a developing flat-plate boundary layer for embedded WMLES. 
Sec.~\ref{sec:conclusion} summarises the paper and draws conclusions.

\section{Derivation of the Hybrid RANS-LES Model}\label{sec:deriv_hrlm}

\subsection{Underlying RSM: SSG/LRR-$\omega$ and -$\ln (\omega)$ RSM}
The RANS part of the present hybrid RANS-LES model is based on a second-moment Reynolds-stress closure.
The underlying differential Reynolds-stress transport equations, which can be directly derived from the Reynolds-averaged momentum equation (cf. Eq.~(\ref{eq:momentum})) is noted in Eq.~(\ref{eq:reynolds_stress_transport_equation}). Here, Reynolds-averaging is applied for density $\Bar{\rho}$ and pressure $\Bar{p}$ and Favre-averaging for the remaining quantities ($\Tilde{\phi}= \frac{\overline{\rho \phi}}{\Bar{\rho}}$):
\begin{align}
\label{eq:reynolds_stress_transport_equation}
    \pdv{( \Bar{\rho}\Tilde{R}_{ij})}{t} 
    + \pdv{\Bar{\rho}\Tilde{R}_{ij}\Tilde{U}_k}{x_k}
    =
    \Bar{\rho}P_{ij}
    +
    \Bar{\rho}\Pi_{ij} 
    - \Bar{\rho}\varepsilon_{ij}
    + \Bar{\rho}D_{ij}
    + \Bar{\rho}M_{ij} \quad .
\end{align}
The terms on the right side are referred to as production term $P_{ij}$, 
pressure-strain correlation $\Pi_{ij}$,
the dissipation term $\varepsilon_{ij}$,
viscous diffusion $D_{ij}$,  
and finally $M_{ij}$ which addresses compressibility effects.
In the present work, the SSG/LRR-$\omega$ RSM is employed, which combines the pressure–strain models of Speziale–Sarkar–Gatski  (SSG) with the Launder–Reece–Rodi model (LRR) close to the wall. 

The required length-scale variable to close the above  equation system is provided by Menter's BSL $\omega$ equation for the original SSG/LRR-$\omega$ RSM \cite{Eisfeld2016}:
\begin{align}
\label{eq:omega_transport_equation}
    \pdv{(\Bar{\rho}\omega)}{t}
    &+
    \pdv{x_k} (\Bar{\rho}\omega \Tilde{U}_k) = \alpha_\omega \frac{\omega}{\Tilde{k}}\frac{\Bar{\rho}P_{kk}}{2} -\beta_{\omega}\Bar{\rho}\omega^2 \\
    &+ \pdv{x_k}\left[\left(     \Bar{\mu} +\sigma_\omega\frac{\Bar{\rho}\Tilde{k}}{\omega}   \right)\pdv{\omega}{x_k}\right]
    +\sigma_d \frac{\Bar{\rho}}{\omega}\max \left( \pdv{\Tilde{k}}{x_k}\pdv{\omega}{x_k},0 \right)
\end{align}
$
$

Alternatively, a transformation of the length-scale variable ($\ln (\omega)$) is used, which typically offers improved numerical robustness without altering the (grid-converged) results of the SSG/LRR-$\omega$ RSM  \cite{eisfeld2022reynolds}. 
On typical grids with limited resolution, however, the numerical discretization might lead to slightly different results of omega- and ln(omega)-based RSM \cite{braun2019implementation}. In order to assess these effects, both variants will be used for all test cases in this paper.
For more details, the reader is referred to the original publications \cite{Eisfeld2016}\cite{eisfeld2022reynolds}.

\subsection{Formulation of the hybridisation approach}
\label{sec:subgrid_models}
The basic idea of DES-based hybrid RANS-LES models is to replace the integral length scale in the corresponding RANS model $l_{RANS}$ by the provided length scale formulation $l_{hyb}$ of the corresponding hybrid method. Along with a sufficient mesh resolution, the adapted model behaves as a Smagorinsky model in LES regions. For the employed SSG-LRR RSM model $l_{RANS}$ is given by:
\begin{align}
 \label{eq:lrans}   l_{RANS} &= \frac{\Tilde{k}^\frac{1}{2}}{c_{\mu}\omega}
\end{align}
with the turbulent kinetic energy $\Tilde{k}$, the specific dissipation $\omega$ and the structure parameter $c_{\mu}$ \cite{Eisfeld2016}. Using the IDDES method, $l_{RANS}$ is replaced by the length scale variable $l_{hyb}$ \cite{shur2008hybrid}:
\begin{align}
\label{eq:lhyb}
l_{hyb} &=  \Tilde{f}_d(1+f_e)l_{RANS} + (1-\Tilde{f}_d)l_{LES} \; .
\end{align}
The function $\Tilde{f}_d = \max \{ (1-f_{dt}), f_B \}$ represents the relevant blending switch between different model branches for Delayed DES (DDES) and wall-modelled LES (WMLES).
The function $f_B$, which is active for the WMLES branch, blends between the RANS mode for the inner turbulent boundary layer ($f_B=1$) and the LES mode for the outer boundary layer ($f_B=0$) and is a purely grid-dependent function (cf. \ref{app:iddes}). 
The purpose of the \emph{elevating} function $f_e$, which is also only active for the WMLES branch, is to prevent a damping of the modelled Reynolds stresses in the area where the RANS and LES modes intersect. Thus, it aims to reduce the commonly observed log-layer mismatch in WMLES.
The LES length scale $l_{LES}$ is defined through the sub-grid length scale $\Delta$ and the calibration constant $c_{DES}$ depending on the corresponding sub-grid model:
\begin{align}
    l_{LES} &= c_{DES} \Delta %
\end{align}

This paper will not focus on the DDES branch of IDDES, which is based on the \emph{RANS shielding} function $f_{d}$.
Instead, we transfer the adjusted formulation of this function from the RSM-DDES in \cite{Probst2011}, i.e.:
\begin{align}
    f_d &= 1-\tanh{\left({16r_d}^3 \right) } \ \ \ \text{and} \\
    \label{eq:eddy_viscosity} r_d &= \frac{\nu + \nu_t}{\sqrt{\sum_{i,j} {\left( \partial u_i/ \partial x_j \right)}^2 } \kappa^2 d^2_w}  \ \ \ \text{with} \ \ \ \nu_t= \frac{\Tilde{k} }{\omega} = \frac{\widetilde{v_i''v_i''} }{ 2 \omega}   \ \ .
\end{align}
The eddy viscosity variable $\nu_t$ is calculated by the modelled turbulent kinetic energy $\Tilde{k}$ and the turbulent length scale $\omega$.
For a complete formulation of IDDES, the reader is referred to \ref{app:iddes} or \cite{shur2008hybrid}.

\subsubsection{Differential hybridisation approach}
In the first considered hybridisation approach proposed in \cite{Probst2011}, 
only the RANS length scale in the dissipation term of the Reynolds-stress equations is replaced by $l_{hyb}$ and no further adjustments are made to the SSG/LRR-RSM model. 
The calibration coefficient in the LES length scale, $c_{DES}=1.1$, is also adopted from \cite{Probst2011}.
Thus, the differential nature of the Reynolds-stress transport equations remains untouched also in the LES regions, where the hybrid model acts as sub-grid scale model.
Therefore, this method is referred to as differential approach in the following.

However, as will be shown in Sec. \ref{sec:dit}, this model does not sufficiently dissipate turbulent kinetic 
energy on the small scales in conjunction with low-dissipation numerical schemes that are commonly used for scale-resolving simulations.
This motivates the formulation of a further hybridisation approach.

\subsubsection{Eddy-viscosity based hybridisation approach}
\label{sec:ev_hyb_approach}
 The insufficient dissipation capability of the differential approach is addressed by switching to an eddy-viscosity-based sub-grid model.
Starting point of the derivation is the compressible Reynolds-averaged momentum equation as formulated in Eq.~(\ref{eq:momentum}):
\begin{equation}
\label{eq:momentum}
\pdv{}{t}\left(\bar{\rho}\tilde{v}_i \right)
+ \pdv{}{x_j}\left(\bar{\rho}\tilde{v}_j \tilde{v}_i \right)
=
- \pdv{\Bar{p}}{x_i}
+\pdv{}{x_j}\left( \Tilde{\tau}_{ij}
- \tau_{ij}^F \right) \; .
\end{equation}

The Favre-averaged Reynolds-stress tensor $\tau_{ij}^F$ of the momentum equation is now approximated by the Boussinesq hypothesis for areas of resolved turbulence, denoted as $\tau_{ij, LES}^F$ (cf. Eq.(\ref{eq:boussinesq})). 
Thus, the components of $\tau_{ij, LES}^F$ are linearly related to respective velocity gradients with the proportionality constant $\mu_t$, denoted as eddy viscosity:

\begin{align}
\label{eq:boussinesq}
    \tau_{ij, LES}^F
    = 
    - \Bar{\rho}\widetilde{v_i''v_j''} 
    \underbrace{=}_{Boussinesq}  %
    2 \mu_T \Tilde{S}_{ij}
    - \left(\frac{2\mu_T}{3}\right)\pdv{\Tilde{v}_k}{x_k}\delta_{ij}  
    - \frac{2}{3}\Bar{\rho}\Tilde{k}\delta_{ij}
\end{align}
The eddy viscosity is calculated by Eq.~(\ref{eq:eddy_viscosity}) and the relation $\mu_t=\Bar{\rho} \nu_t$. 
For this calculation, the turbulent kinetic energy $\Tilde{k}$ and the turbulent length scale $\omega$ quantities are directly provided by the original SSG/LRR-RSM equations.
The dissipation capability of this sub-grid model is successfully demonstrated in Sec. \ref{sec:dit}, and the $c_{DES}$ constant is calibrated to 0.65.
In this work we consider two different approaches to integrate the eddy-viscosity sub-grid model into the IDDES method:

\paragraph{\textbf{Integration of sub-grid model into HRLM: Variant (a)}}
In this variant, the eddy-viscosity-based approximation of the 
Reynolds stress tensor $\tau_{ij, LES}^F$ is only applied in LES regions, while all RANS regions (including the near-wall layer in WMLES) should retain the differential RSM nature (denoted as $\tau^F_{ij, RANS}$).
To achieve this, a suitable transition between these areas is required which is given by the blending function $\Tilde{f}_d$ of the IDDES method (cf. Eq.~(\ref{eq:lhyb})): %
\begin{equation}
\label{eq:blending_rst}
    \tau^F_{ij, hyb} = \Tilde{f}_d \tau^F_{ij, RANS}  +  (1-\Tilde{f}_d)\tau^F_{ij, LES}
\end{equation}
Additionally, the choice of the IDDES model constants $c_l$ and $c_t$
of the elevating function $f_e$ in Eq.~(\ref{eq:lhyb}) are investigated in Sec. \ref{sec:channel_variant_a}. 
The best results are achieved for values of $c_l=5$ and $c_t=1.87$, which correspond to the values chosen in the SST-based IDDES.

\paragraph{\textbf{Integration of sub-grid model into HRLM: Variant (b)}}

In this model variant the blending function $\Tilde{f}_d$ of Eq.~(\ref{eq:blending_rst}) is replaced by a newly introduced binary function $\Tilde{f}_{user}$.
This function is defined to achieve 
the discrete values 1 or 0 depending on a manually specification.
This allows for a more flexible application of the Boussinesq hypothesis on the Reynolds Stress tensor.

\begin{equation}
\label{eq:blending_rst_user}
    \tau^F_{ij, hyb} = \Tilde{f}_{user} \tau^F_{ij, RANS}  +  (1-\Tilde{f}_{user})\tau^F_{ij, LES}
\end{equation}

In the present work, we investigate flows with the following definition of $\Tilde{f}_{user}$:
\begin{itemize}
    \item (WM)LES regions including near-wall RANS layers: $\Tilde{f}_{user} = 0$
    \item RANS regions outside embedded WMLES domains: $\Tilde{f}_{user} = 1$
\end{itemize} 

The choice of $\Tilde{f}_{user} = 0$ implies that the underlying SSG/LRR-RSM is locally modified and reduced to an eddy-viscosity model, also in the near-wall RANS layer of WMLES. 
Thus, in contrast to model variant (a), no blending between near-wall RANS layers
and corresponding WMLES regions is employed, but both regions are treated with $\tau^F_{ij, LES}$.
However, the differential character of the RSM is not completely vanished but implicitly appears in the quantities of $\Tilde{k}$ and $\mu_T$. This is because these values are still calculated according to the Reynolds Stress transport equation (cf. Eq.~(\ref{eq:reynolds_stress_transport_equation}) and (\ref{eq:eddy_viscosity})).

As in variant (a) the IDDES coefficients $c_l$ and $c_t$ are calibrated in a periodic channel flow (cf. Sec. \ref{sec:channel_variant_a}). 
Again, the best performance can be achieved for the values $c_l=5$ and $c_t=1.87$.

\section{Numerical Method}\label{sec:num_meth}
The present work employs an unstructured compressible finite-volume solver, the DLR-TAU code \cite{Schwamborn2006}, for solving the corresponding flow and model equations. 
Even though only hexahedral grids are used throughout this work, the solver treats these cells in the same unstructured manner as e.g. tetrahedra or prisms using a vertex-centred dual-cell approach.
The discretization schemes for both space and time are basically of 2nd order.
To fulfil the commonly-accepted accuracy requirements for scale-resolving simulations, in particular when applied to wall-bounded flow, we employ a low-dissipation low-dispersion (LD2) spatial scheme, which is briefly outlined in Sec.~\ref{sec:ld2}.
On the temporal side, a 2nd-order implicit backward-difference scheme is solved using dual-time stepping with sub-iterations and Cauchy-convergence criteria for relevant local and integral flow quantities.

When applying hybrid RANS-LES to rather stable wall-bounded flows like the flat-plate boundary layer in Sec.~\ref{sec:flat_plate}, the transition from RANS modelling to (wall-modelled) LES needs to be augmented in order to avoid large "grey areas". 
To this end, we employ an embedded WMLES approach based on local synthetic-turbulence injection, which is briefly outlined in Sec.~\ref{sec:stg}.

\subsection{Low-dissipation low-dispersion scheme}\label{sec:ld2}
Like most other hybrid RANS-LES models, the original IDDES \cite{shur2008hybrid} has been designed and calibrated under the assumption that all physical dissipation of small-scale turbulence in the LES regions is provided by the sub-grid scale model, whereas numerical errors from the discretization scheme are assumed negligible.
Consequently, applying the IDDES in combination with rather high (e.g. artificial) numerical dissipation as typically used for RANS computations can lead to significant errors in the prediction of mean-flow quantities, as e.g. shown in SST-IDDES computations of a periodic channel flow in \cite{Carlsson2023}.   
In this respect, we not only consider the spectral turbulent energy decay (mainly related to the sub-grid model constant $C_{DES}$) to be sensitive to the numerical scheme, but also the wall-normal transition from RANS to LES in the WMLES mode of IDDES, which depends on the well-defined interaction of the various IDDES functions (cf. Sec.~\ref{sec:subgrid_models}).
Therefore, a discretization scheme providing sufficiently low numerical errors is considered crucial for accurate flow predictions using IDDES, including the new RSM-based variant introduced in Sec.~\ref{sec:subgrid_models}.

A suitable numerical scheme, which provides both low dissipation and dispersion errors for unstructured finite-volume solvers is the \emph{LD2 scheme} according to \cite{Loewe2016}, \cite{Probst2016}.
The scheme is based on a skew-symmetric central flux for the convection terms according to Kok \cite{kok2009high} which provides local and global energy conservation for incompressible flow on curvilinear grids. 
To ensure stability for compressible flow and general (unstructured) grids, matrix-valued artificial dissipation is added which, however, is strongly reduced compared to standard central schemes.
Most importantly, the global 4th-order dissipation coefficient takes values as small as $\kappa^{(4)} = 1/1024$ compared to $\kappa^{(4)} = 1/64$ typically used for RANS computations with the TAU solver.
Additionally, a low-Mach-number preconditioning matrix is introduced in the dissipation operator to consistently scale the dissipation towards the incompressible limit.
Details of the manual optimisation procedure of the artificial-dissipation parameters to arrive at a sufficiently low dissipative scheme for scale-resolving simulations, called \emph{LD scheme}, are found in \cite{Probst2015a}.

Besides dissipation, the complete \emph{LD2 scheme} also addresses the rather large dispersion errors of standard central schemes.
To this end, the left and right face values $\phi_{L, ij}$, $\phi_{R, ij}$ entering the skew-symmetric fluxes are linearly extrapolated using the local Green-Gauss gradients $\nabla_0 { \phi }$, similar to a MUSCL-type reconstruction.
For illustration, the extrapolated flux term for a generic variable $\phi$ reads:
\begin{equation}\label{eqn:ld2_flux}
\phi_{ij, \alpha_e} = \frac{1}{2} \left( \phi_{L, ij} + \phi_{R, ij} \right) =  \frac{1}{2} \left( \phi_i + \phi_j \right) + \frac{1}{2} \alpha_e \left( \nabla_0  { \phi_i} - \nabla_0 { \phi_j } \right) \cdot \mathbf{d}_{ij}  \quad .
\end{equation}

By applying such a reconstruction to the 2nd-order central scheme, we gain an additional degree of freedom via the extrapolation parameter $\alpha_e$ which could be used to increase the accuracy order on sufficiently smooth grids (i.e. where simple Green-Gauss gradients are accurate), cf. \cite{Loewe2016}.
However, since we consider the actual error properties of the scheme for a given grid resolution to be more important than the asymptotic order of accuracy, $\alpha_e$ has been used to minimise the dispersion error based on a generic 1-D wave convection problem.
With the determined value of $\alpha_e = 0.36$, the required number of points to achieve given dispersion error levels could be reduced more than with a standard 4th-order discretization, cf. \cite{Loewe2016}.

The combination of the low-dissipation flux with low-dispersive extrapolation forms the \emph{LD2 scheme}, which is used throughout this work.
In \cite{Probst2016} it has been shown to further improve the accuracy of IDDES in periodic channel flows compared to the basic low-dissipation (LD) scheme. 
Recently, the scheme was implemented and applied in another compressible flow solver and showed similar accuracy improvements in IDDES computations compared to a standard central scheme \cite{Carlsson2023}.

\subsection{Embedded WMLES with Synthetic Turbulence Injection}\label{sec:stg}
The validation of the WMLES branch for developing boundary layers performed in Sec.~\ref{sec:flat_plate} requires a method to rapidly switch from RANS to WMLES mode in the streamwise direction.
Since no separation is involved, the RANS-LES switching location is prescribed manually by setting $f_{dt} = 1$ in the WMLES region (cf. Eq.~(\ref{eq:lhyb})),
and synthetic turbulence is injected near the interface to augment rapid development of resolved turbulence.
This approach, denoted as embedded WMLES, applies the Synthetic Turbulence Generator (STG) of Adamian and Travin \cite{Adamian2011a} with volumetric-forcing extensions by Francois \cite{francois2015forced}.
This STG computes local synthetic velocity fluctuations by superimposing a set of $N$ Fourier modes as:
\begin{equation}\label{eq:STG_fluc}
 \vec{u}'_{ST} 
 = \vec{\underline{A}} \cdot \sqrt{6}\sum_{n=1}^N\sqrt{q^n}\left[\vec{\sigma}^n\cos\left(k^n\vec{d^n}\cdot\vec{r}'+\phi^n+s^n\frac{t'}{\tau}\right)\right] \quad .
\end{equation}
Random values, which are fixed throughout the simulation, are generated to determine stochastic distributions of the direction vectors $\vec{d}^n$ and $\vec{\sigma}^n \perp \vec{d}^n$, the mode phase $\phi^n$, and the mode frequency $s^n$.
The mode amplitudes $q^n$ are derived from a von K\'{a}rm\'{a}n model spectrum for the turbulent kinetic energy, which is constructed from RANS-input data and local grid properties. 
These grid properties consist of local maximal cell widths defining a function $f_{cut}$, which cuts off the von K\'{a}rm\'{a}n model spectrum above the Nyquist wave number.
An important model parameter that controls the size of the large energy-containing synthetic fluctuations is the length scale $l_e$:
\begin{equation}\label{eq:STG_le}
 l_e = \min \left( 2 d_w, C_l l_t \right) \quad \mbox{with:} \;\; C_l = 3 \text{ and } l_t = \frac{{\left(\frac{1}{2}  \overline{u_i'u_i'} \right)}^{3/2}}{\varepsilon}   
\end{equation}

which depends on the wall distance $d_w$ and the integral length scale $l_t$ from RANS-input data.
The integral length scale $l_t$ is defined by the turbulent kinetic energy $(\frac{1}{2} \overline{u_i'u_i'})$ as well as the dissipation rate $\varepsilon$.
Note that even though the value of $C_l$ was determined based on a plane mixing layer  \cite{Shur2014}, this STG has proven to yield convincing results in wall-bounded flows as well, such as periodic channels and flat plates \cite{Shur2014}, \cite{Probst2020}.
The input data is extracted from upstream of the RANS-LES interface and is also used to scale the synthetic fluctuations via the Cholesky-decomposition of the Reynolds-stress tensor $\vec{\underline{A}}$.

Instead of abrupt synthetic forcing in the interface plane, we employ a smoothly varying source-term forcing in a small volumetric domain (typically spanning half a local boundary-layer thickness) downstream of the interface. 
To ensure realistic temporal correlations of the synthetic fluctuations when travelling through this domain, the position vector $\vec{r'}$ and the time $t'$ are modified in accordance with Taylor's frozen velocity hypothesis, cf. \cite{francois2015forced}.

The forcing source term approximates the partial time derivative of the synthetic velocity fluctuations consistently with TAU's $2^{nd}$-order backward time-discretization scheme:
\begin{equation}\label{eq:forcing_source}
\vec { Q } = \frac{\partial\left(\rho \vec{u}'_{ST}\right) }{\partial t}  \approx \frac { 3\left( {\rho \vec{u}'_{ST} } - \rho \vec{u}'^n \right) -\left( \rho \vec{u}'^n - \rho \vec{u}'^{n-1} \right)  }{ 2\Delta t }  \ .
\end{equation}
By using running time averages to compute the actual "previous" fluctuations, i.e. $\vec {u' } ^{ n } = \vec {u} ^{ n } -  \langle \vec {u} \rangle $ and $\vec { u' } ^{ n-1 } = \vec {u} ^{ n-1 } -  \langle \vec {u} \rangle $, a rather accurate reproduction of the target fluctuations is achieved.
For the desired smooth injection, a blending factor $\in [0,1]$ with Gauss-like distribution scales the source within the forcing domain, see \cite{Probst2020} for details.

\section{Validation Results}\label{sec:results}

\subsection{LES functionality}
\label{sec:dit}
Hybrid RANS-LES methods based on Detached-Eddy simulation (DES) are designed to function as Smagorinsky-type models in LES regions. This is achieved through the adaption of the integral length scale in the  underlying RANS model from $l_{RANS}$ to $l_{LES}$ in LES regions (cf. Sec. \ref{sec:subgrid_models}).
The coefficient $c_{DES}$ within $l_{LES}$ has to be chosen such that the level of modelled turbulence of the new eddy-viscosity-based SSG/LRR sub-grid model is similar to a Smagorinsky-type sub-grid model. Therefore, the widely used approach to empirically calibrate the $c_{DES}$ constant using experimental reference data for decaying isotropic turbulence (DIT) from \cite{comte1971simple} is followed in this work. The reference data consists of kinetic energy spectra of isotropic grid turbulence at different time levels $\left( t \in \{0\,\text{s}, 0.87\,\text{s}, 2\,\text{s}\}\right)$
A suitable test case setup for DIT is adopted from \cite{Probst2011}. This setup comprises a cubic computational volume with an edge length of $2\pi$. Periodic boundary conditions are applied to opposing surfaces in every spatial direction.
To validate the calibration of $c_{DES}$ for a range of spatial resolutions, the computational domain is discretised by three different isotropic resolutions $\left( \Delta x  \in \{\frac{2\pi}{32} , \frac{2\pi}{64} , \frac{2\pi}{128} \}\right)$. Suitable temporal discretisations were derived from time-step convergence studies and lead to time step sizes of $\Delta t=0.005\,$s for the coarse and medium grid ($32^3$ and $64^3$ cells) and $\Delta t=0.002\,$s for the finest grid ($128^3 $ cells).

\begin{figure}
	\begin{center}
	\begin{subfigure}[c]{0.39\textwidth}
			\includegraphics[trim= 250 150 50 90, clip,width=\linewidth]{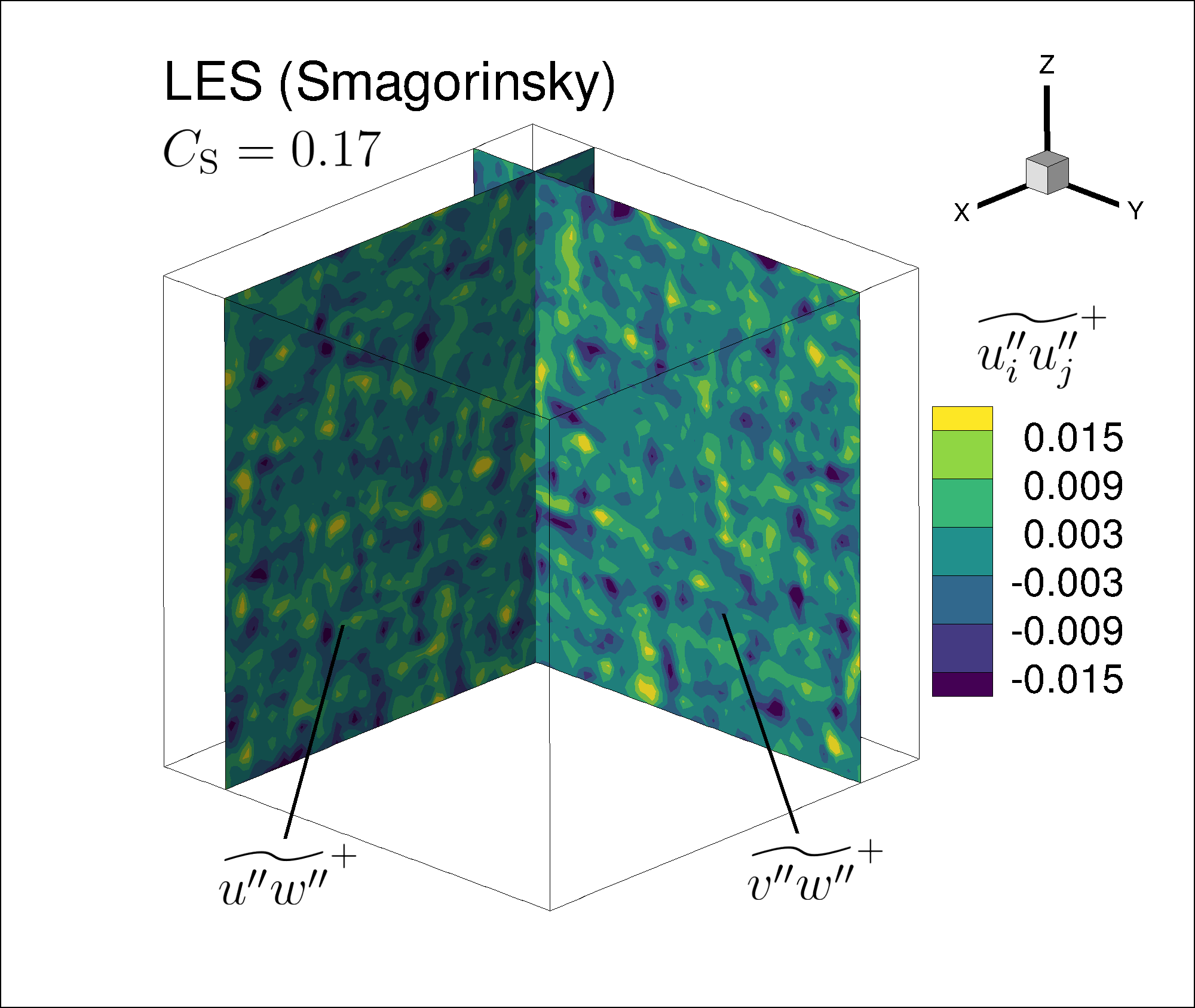}	
	       	\caption{}
    \end{subfigure}	%
 	\begin{subfigure}[c]{0.29\textwidth}
			\includegraphics[trim= 250 150 450 90, clip,width=\linewidth]{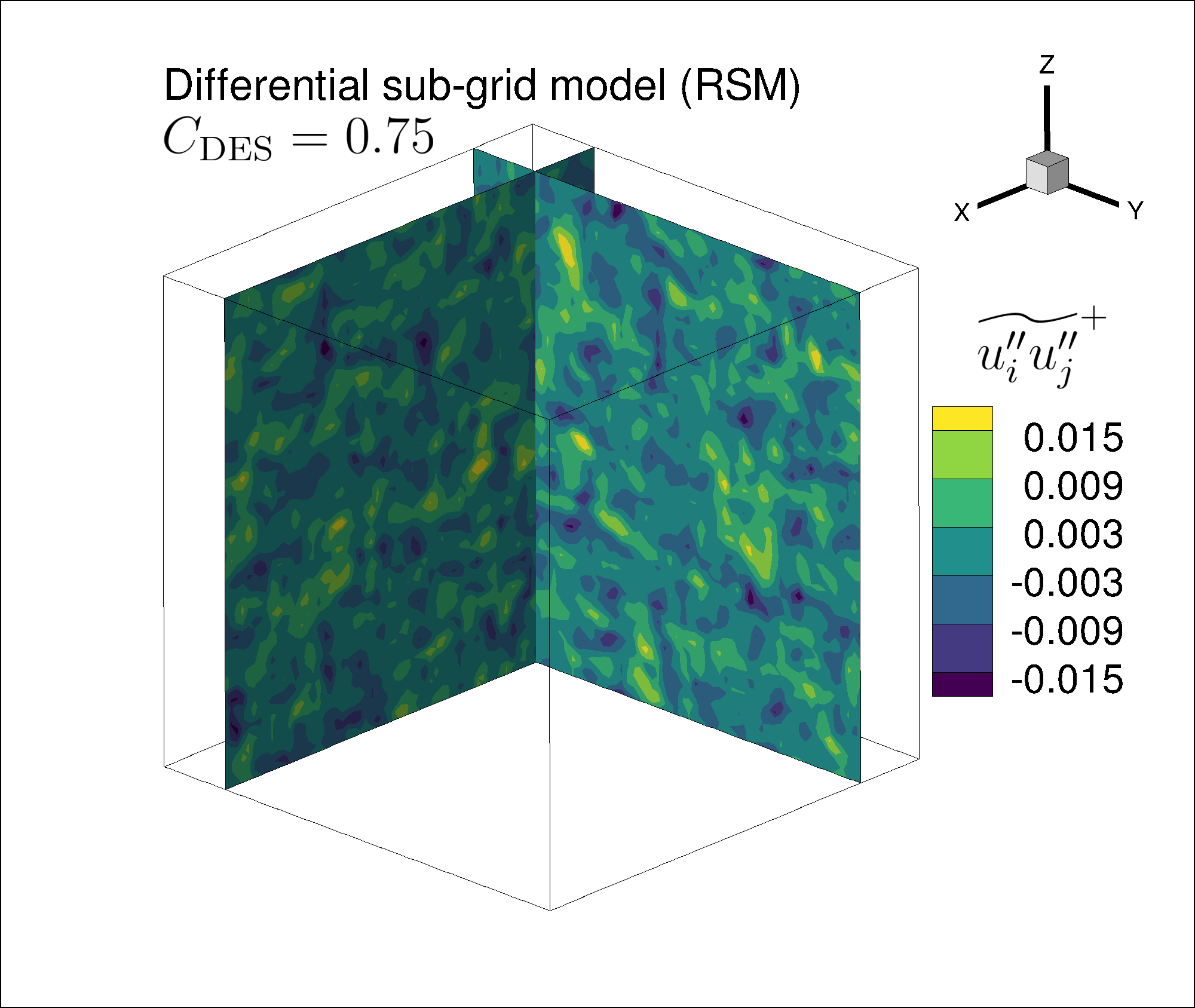}	
	       	\caption{}
    \end{subfigure}	%
	\begin{subfigure}[c]{0.29\textwidth}
			\includegraphics[trim= 250 150 440 90, clip,width=\linewidth]{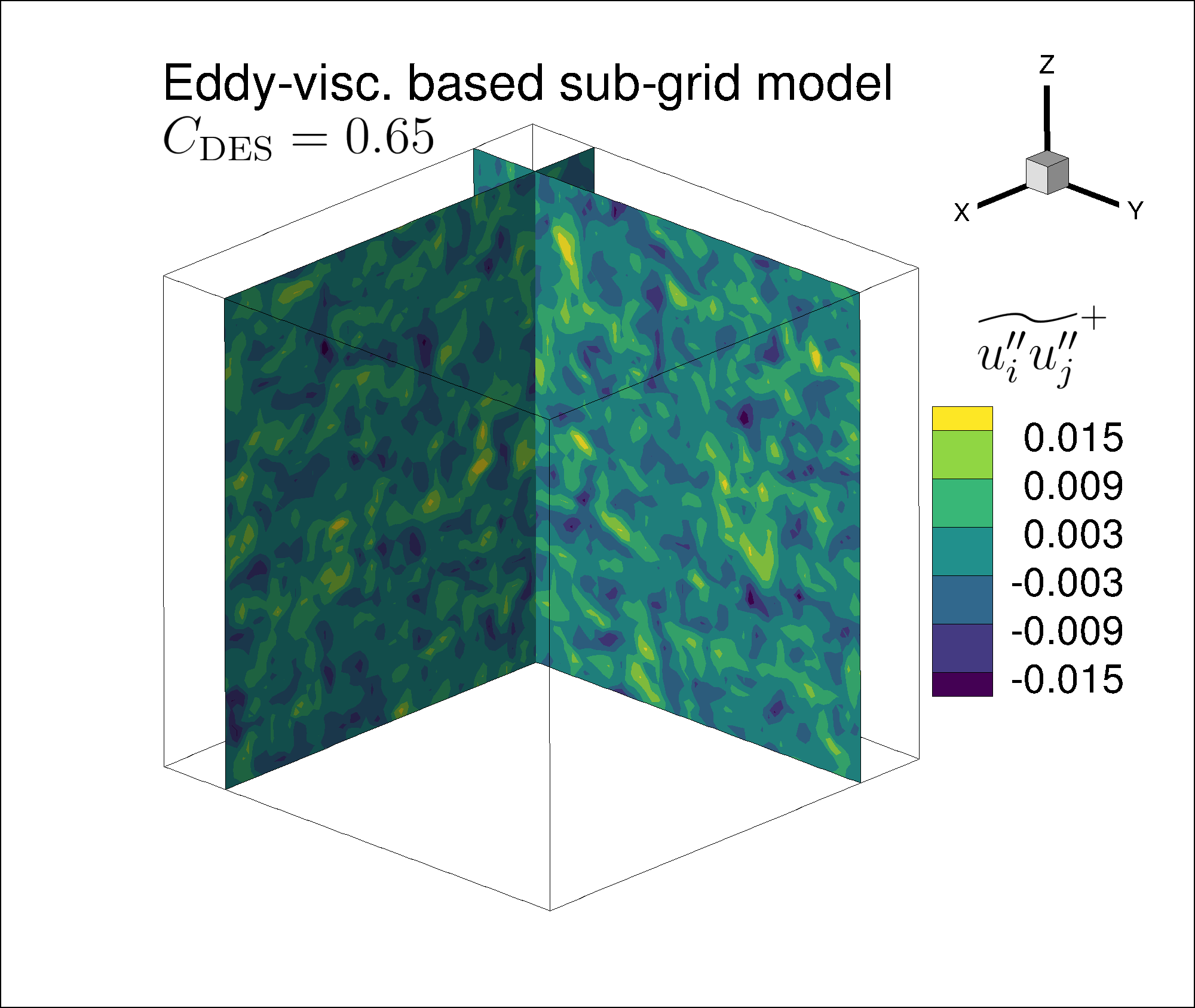}	
	       	\caption{}
    \end{subfigure}	%
	\caption{Initial flow fields of the DIT (Decaying Isotropic Turbulence) test case at $t=0\,s$ and $64^3$ grid cells. Selected modelled Reynolds shear stresses are presented for different sub-grid models. The stresses are normalised with the variance of the velocity field.}
	\label{abb:DIT_initial}
	\end{center}
\end{figure}

\begin{figure}
\begin{center}
\begin{subfigure}[c]{0.48\textwidth}
			\includegraphics[trim= 90 90 180 180, clip,width=\linewidth]{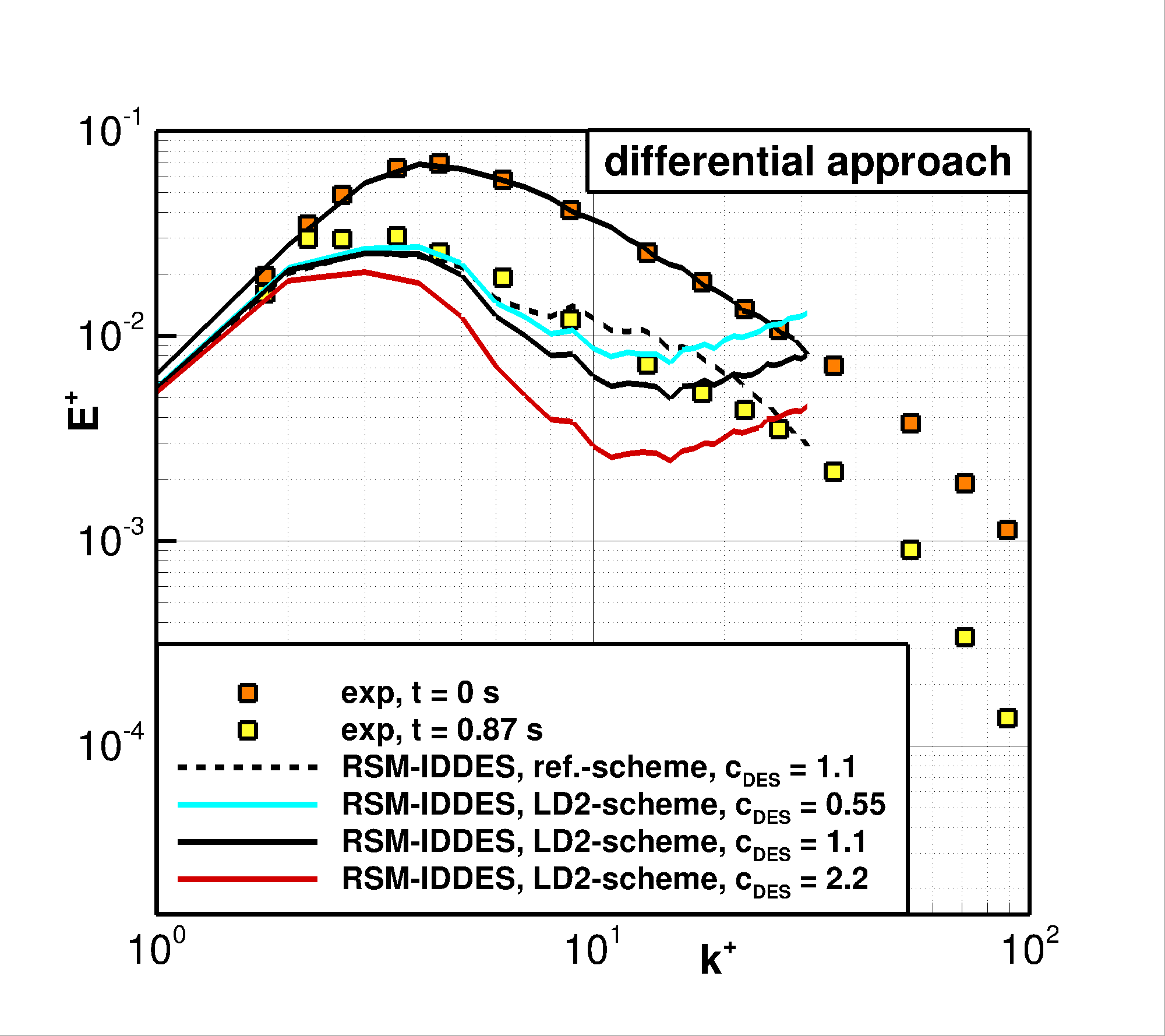}	
            \caption{
            Results of the differential RSM sub-grid model for different $c_{\text{DES}}$ constants and numerical schemes.}
            \label{abb:dit_approaches}
\end{subfigure}  \hspace*{0.3cm}
 \begin{subfigure}[c]{0.48\textwidth}
			\includegraphics[trim= 90 90 180 180, clip,width=\linewidth]{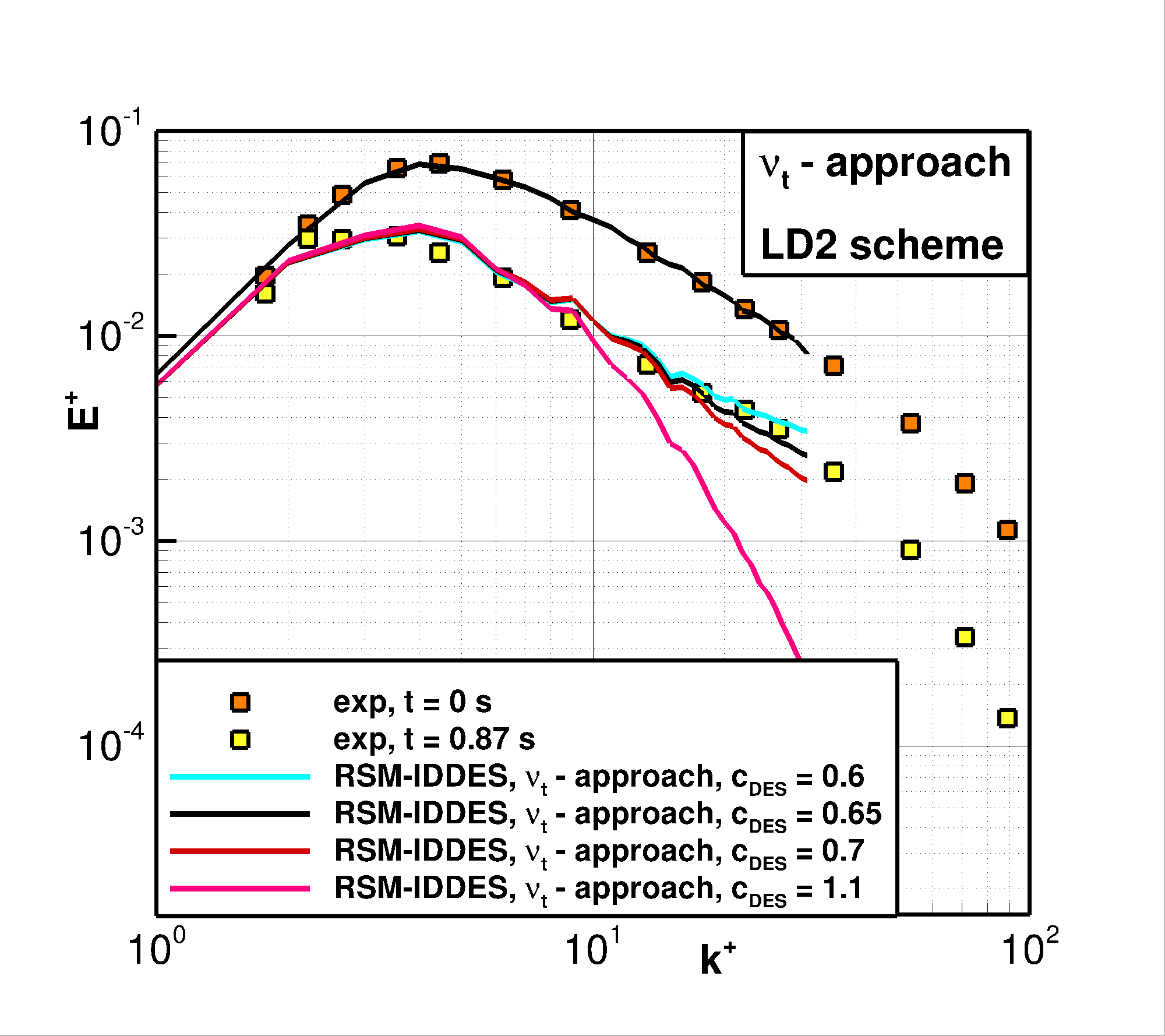}	
	        \caption{Calibration of the $c_{DES}$ constant for the eddy viscosity based RSM sub-grid model with the LD2 scheme.}
            \label{abb:dit_64_calib}
\end{subfigure}%
\caption{Turbulent kinetic energy spectra of decaying isotropic turbulence (DIT) at two times ($t=0\,s; t=0.87\,s$). The performance of two different LES sub-grid models is assessed by experimental reference data \cite{comte1971simple}.}
\label{abb:dit_main_results}
\end{center}
\end{figure}

\paragraph{Initialisation of the computational setup}
The setup is initialised by a velocity field calculated from the experimental energy spectra at $t=0\,\text{s}$. To this end, the energy spectra is transformed to local space by a method inspired by \cite{Kraichnan1970}. Due to the compressible flow solver, additional density and pressure fields are required and calculated from the generated velocity field using the compressible Bernoulli equations.
Additionally, the modelled Reynolds-stress tensor and the specific dissipation rate $\omega$ ($\ln (\omega)$) are computed in a stationary RSM-IDDES calculation. In this calculation the previously initialised fields are kept constant and only the turbulence equation are solved so that the modelled Reynolds-stresses can adapt to the prescribed data. 
Selected initial solutions obtained with the initialisation approach are presented below. With a suitable choice of the respective $c_{DES}$ constants, shear stresses of the novel RSM-IDDES hybridisation approaches (differential a eddy viscosity based RSM) match the Smagorinsky-LES reference initialisation (Fig. \ref{abb:DIT_initial}). Furthermore, direct agreements of the corresponding energy spectra at $t=0\,s$ with the experimental reference are obtained (Fig. \ref{abb:dit_main_results}). %
In the following, several scale-resolving RSM-IDDES simulations are applied to the computational setup and the actual decay of the initial turbulence is calculated over a total simulation time of $2\,$s.

\paragraph{Results of differential approach}
Initially, results of the differential sub-grid model as introduced in Sec. \ref{sec:subgrid_models} are presented. 
Fig. \ref{abb:dit_approaches} shows different normalised energy spectra $E^+$ in dependence of the normalised wave number $k^+$ at a time level of $t=0.87\,$s together with experimental data.
First, consider results from using a numerical scheme with rather high numerical dissipation, using a 
4th-order dissipation coefficient of $k^{(4)} = 1/128$.
The combination of this high-dissipation scheme with the differential sub-grid model shows rather good agreement with experimental data (black dashed curve in \ref{abb:dit_approaches}).
These results are consistent with findings in \cite{Probst2011} where the differential approach was originally introduced to couple an RSM with DES, also using a rather dissipative numerical scheme. 

However, moving to the low-dissipative LD2-scheme that was found suitable for more complex DES simulations \cite{Probst2016}, \cite{Carlsson2023}, considerable deviations appear. 
The turbulent kinetic energy (TKE) is not sufficiently dissipated by the differential sub-grid model which leads to an accumulation of TKE in the small turbulent scales. 
As Fig. \ref{abb:dit_approaches} demonstrates, this accumulation of TKE cannot be circumvented by adjusting the $c_{DES}$ constant. For all selected values of $c_{DES}$ significant deviations appear and no agreement of the overall energy spectrum with the reference data can be achieved.
The effect of energy accumulation can be explained by the missing linear alignment of the modelled stress tensor from the local strain-rate tensor, which is strictly enforced in common eddy-viscosity-based sub-grid models like Smagorinsky, WALE or SA- and SST-based DES.
In general, momentum terms with damping or dissipative properties, such as physical diffusion or (artificial) numerical dissipation, are directly proportional to differences of gradients of the velocity field. However, the Reynolds stresses of the differential RSM-IDDES are only indirectly coupled with velocity gradients, e.g. via the production and redistribution terms.
In other words, one of the main features of differential Reynolds-stress models that allow for accurate RANS modelling of complex phenomena (e.g. vortices or streamline curvature) apparently prevents this type of model to act sufficiently dissipative when used as DES sub-grid model, unless supported by rather high numerical dissipation.

\begin{figure}
	\begin{center}
	\begin{subfigure}[c]{0.32\textwidth}
			\includegraphics[trim= 90 90 180 180, clip,width=\linewidth]{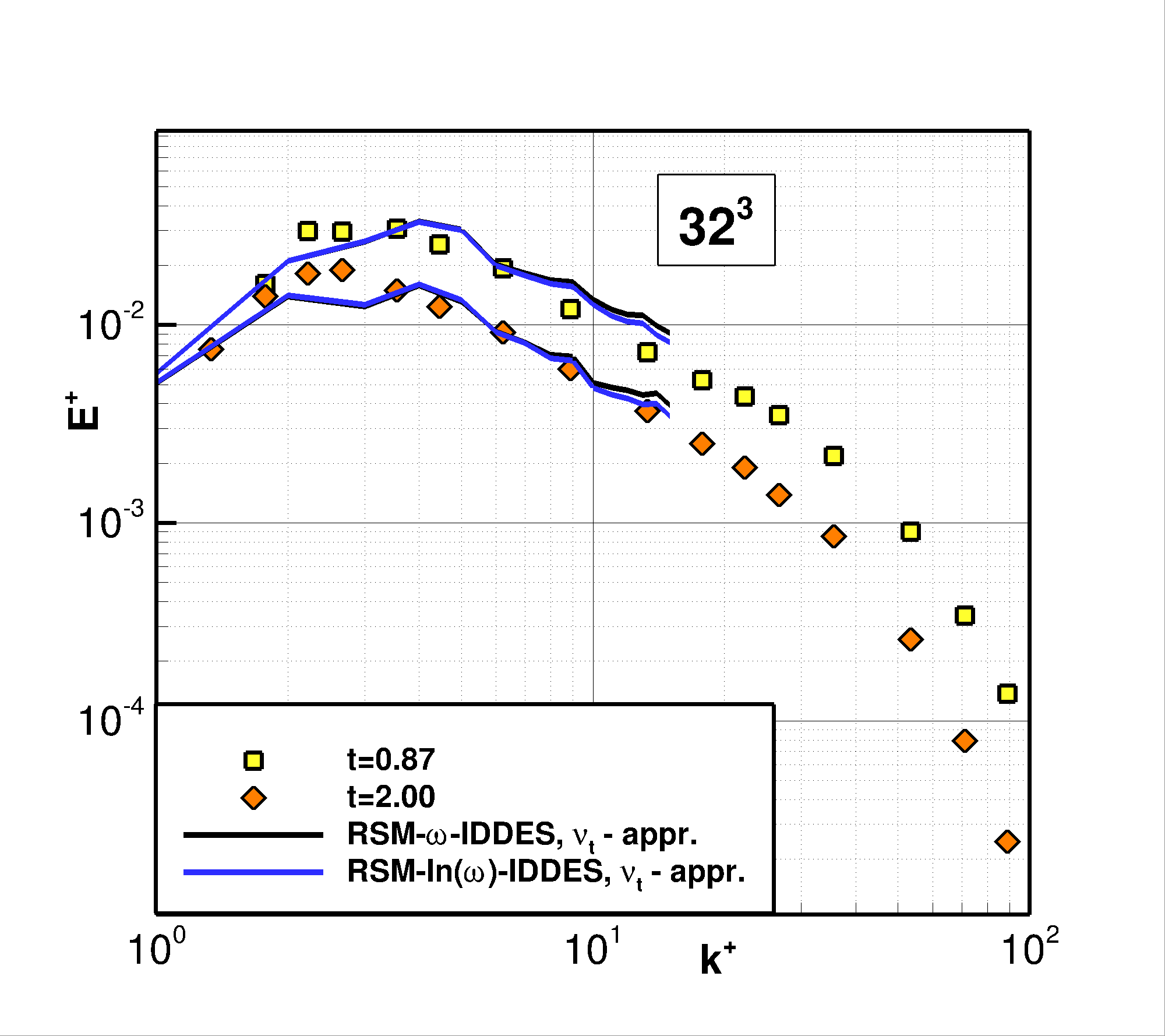}	
	       	\caption{}
	        \label{abb:DIT_resolution_32}
    \end{subfigure}	%
 	\begin{subfigure}[c]{0.32\textwidth}
			\includegraphics[trim= 90 90 180 180, clip,width=\linewidth]{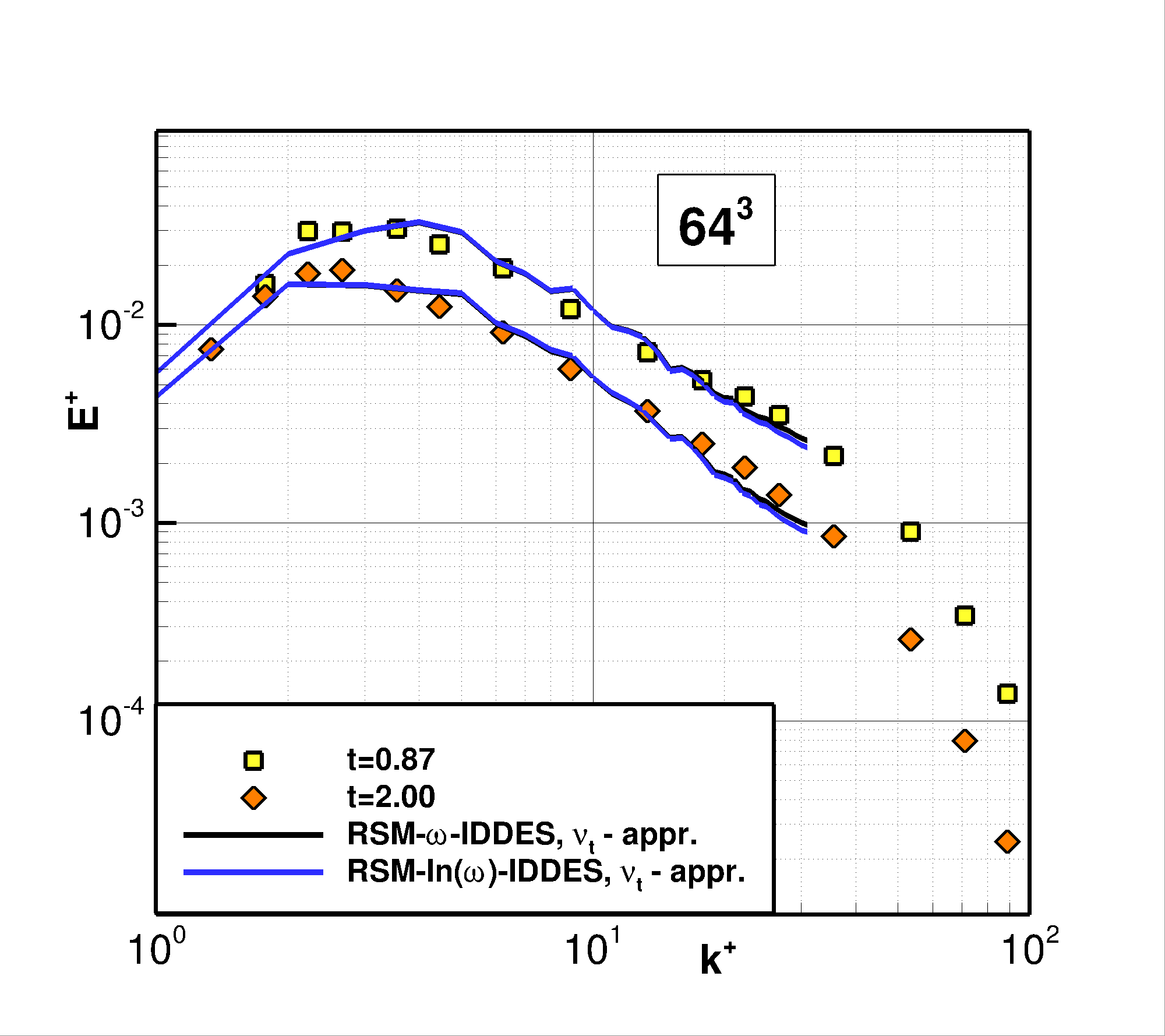}	
	      	\caption{}
	        \label{abb:DIT_resolution_64}
    \end{subfigure} %
	\begin{subfigure}[c]{0.32\textwidth}
			\includegraphics[trim= 90 90 180 180, clip,width=\linewidth]{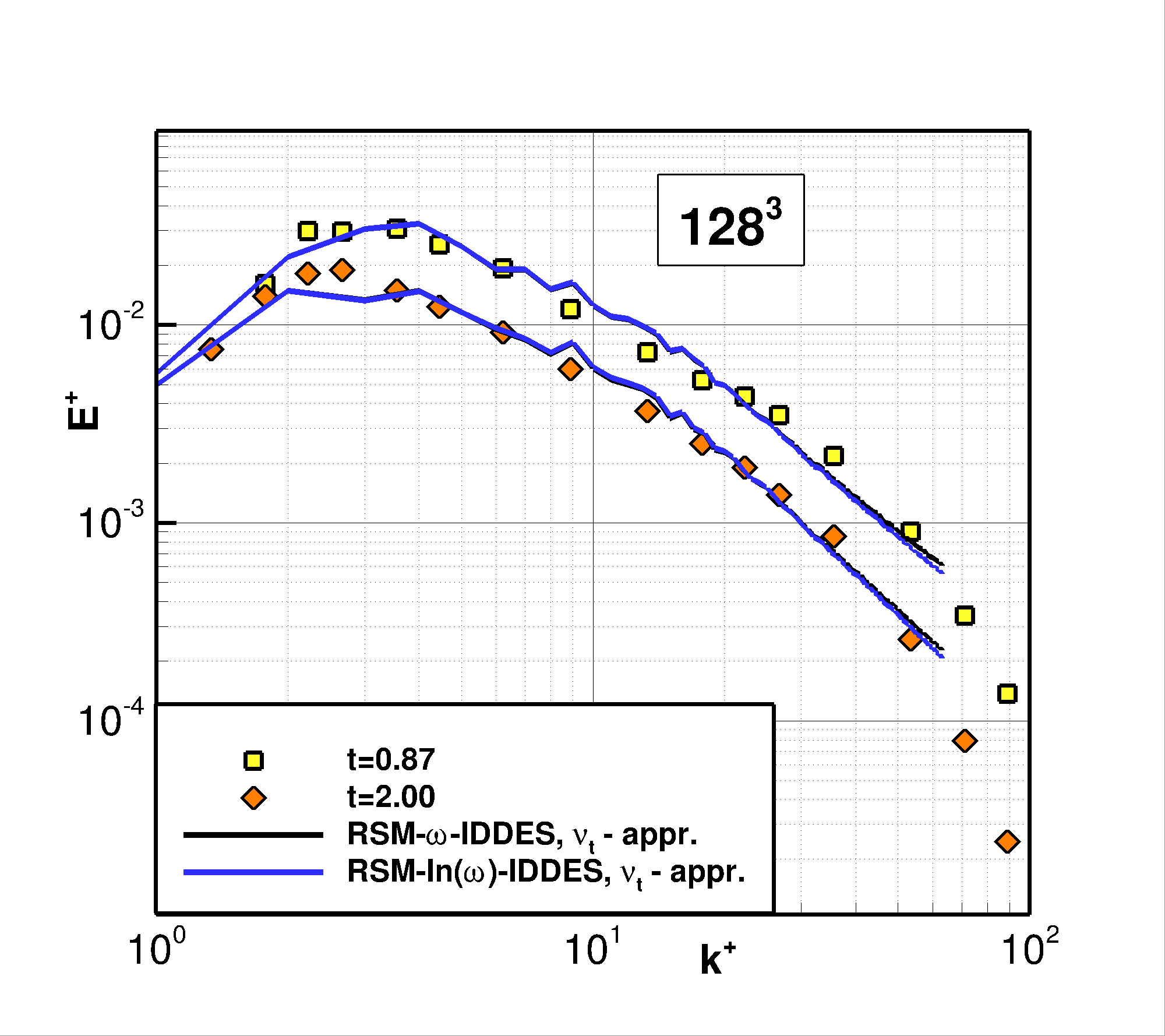}	
        	\caption{}
	        \label{abb:DIT_resolution_128}
	\end{subfigure}%
	\caption{Energy spectra of RSM-$\omega$-IDDES and RSM-$\ln (\omega)$-IDDES with calibrated $c_{DES}=0.65$ constant for different spatial resolutions and times ($t=0.87\,s; t=2\,s$).}
	\label{abb:DIT_resolution}
	\end{center}
\end{figure}

\paragraph{Results of eddy-viscosity based RSM sub-grid model}
In the following, calibration results of the eddy-viscosity-based sub-grid model (cf. Sec.~ \ref{sec:subgrid_models}) in conjunction with the LD2 scheme are presented.
The initial value of the $c_{DES}$ calibration constant was set to 1.1, which corresponds to the value of the differential sub-grid model in \cite{Probst2011}. 
As Fig. \ref{abb:dit_64_calib} shows, good agreements are obtained for low wave numbers ($k^+ \leq 10$) whereas the energy is underestimated for higher wave numbers ($k^+ \geq 10$).
Since the $c_{DES}$ constants correlates to the level of eddy viscosity which in turn impacts the dissipation of turbulent kinetic energy, one can conclude that the initial value of $c_{DES}=1.1$ is too high. Further RSM-IDDES simulations with reduced $c_{DES}$ constants $\left( c_{DES} \in \{0.6, 0.65, 0.7\}\right)$ lead to an significantly enlarged energy content of the small turbulent scales (high wave numbers) and therefore better agreements with the experimental spectrum (cf. Fig \ref{abb:dit_64_calib}).
Considering the agreement between the numerical and experimental results for both time levels and all spatial resolutions, an optimal value of $c_{DES}=0.65$ was derived (cf. Fig. \ref{abb:DIT_resolution_32}-\ref{abb:DIT_resolution_128}). 
Additionally, further simulations were performed with RSM-IDDES and a logarithmic length scale variable ($\ln (\omega)$) in Menter's BSL $\omega$ equation (denoted as RSM-$\ln (\omega ) $-IDDES)  (cf. Fig. \ref{abb:DIT_resolution_32}-\ref{abb:DIT_resolution_128})\cite{eisfeld2022reynolds}. Again, very good agreement with the experimental spectra can be seen. Minor deviations between the $\omega$ and $\ln ( \omega )$ formulation of the RSM-IDDES only appear with respect to the higher wave numbers $k^+$. However, since the integral length scale $l_{RANS}$ is entirely replaced by $l_{LES}$ for this test case, no differences between both RSM formulations are expected. A potential reason for the minor deviations may be slightly different flow-field initialisations.

\subsection{WMLES functionality}
\label{sec:channel}
After the demonstration of the LES functionality of the newly developed RSM-IDDES, the wall-modelled LES (WMLES) branch of IDDES is addressed in this section. 
Therefore, wall-bounded flows in form of fully developed turbulent channel flows are simulated and serve as suitable test case for the WMLES validation.
Note that the WMLES branch of IDDES is enforced for this test case by setting the IDDES model function $f_{dt}$ to 1. Thus, $\Tilde{f_d}$ from Eq. \ref{eq:fdtilde} simplifies to the WMLES blending function $f_B$. Although the IDDES method would automatically switch to WMLES mode for this test case, this procedure ensures that only the WMLES branch is examined.

In this work the computational setup and flow conditions are adapted from \cite{probst2016scale}.
The computational domain consists of a rectangular cuboid with height $2h$, length $2\pi h$ and width $\pi h$, where $h$ is defined as the channel half height. For the fully developed turbulent flow, the half height $h$ directly corresponds to the boundary layer thickness $\delta$.
Periodic boundary conditions are applied in streamwise as well as spanwise directions whereas no-slip wall conditions are present for the upper and lower face. 

To ensure a constant bulk velocity $U_{bulk}$ throughout the simulation an additional pressure gradient is applied in streamwise direction. This counteracts the natural decrease of bulk velocity which would arise otherwise due to viscous dissipation.
Channel flows at two different bulk Reynolds numbers $Re_{\delta} =U_{bulk}\cdot\delta/\nu$ are considered, $Re_{\delta} \in \{6875, 98300 \}$, which allows for comparisons with DNS reference data with identical values of $Re_{\delta}$. \cite{moser1999direct} \cite{lozano2014effect}.

\paragraph{Spatial and temporal resolution}
With regard to the spatial discretization a fully structured mesh with 65 hexahedral cells in streamwise ($x$) and spanwise ($z$) directions is employed and results into resolutions of 
$\delta / \Delta x= 10$ and $\delta / \Delta z= 20$.
The spatial resolution in wall-normal ($y$) direction is characterised by a wall-adjacent grid spacing of $y^+(1)=0.2$ and a growth rate of $r=1.14$ leading to 65 cells for $Re_\delta = 6875$ and 141 cells for $Re_\delta = 98300$. 
The grid spacing of $y^+(1)=0.2$ safely fulfils the resolution requirement of the RSM-RANS model, which is $\Delta y^+ \leq 0.4$ for the selected numerical setup in the DLR-TAU code.
For both flow simulations a normalised time step size of $\Delta t^+ = 0.4$ is selected which fulfils $CFL\leq 1$ in the entire computational domain. 

The channel flow is simulated for $30$ convective time units (CTUs) where a single CTU is defined as  
$\text{CTU} = 2\pi \delta / U_{bulk}$.
During the initial $10\,$CTUs, the flow evolves from an initial RSM-RANS solution to a fully developed 
WMLES solution.
For the actual results, only the last $20\,$CTUs are used to calculate selected statistical flow quantities. 
\subsubsection{RANS results}
\begin{figure}
\begin{center}
\begin{subfigure}[c]{0.48\textwidth}
			\includegraphics[trim= 110 90 100 80, clip,width=\linewidth]{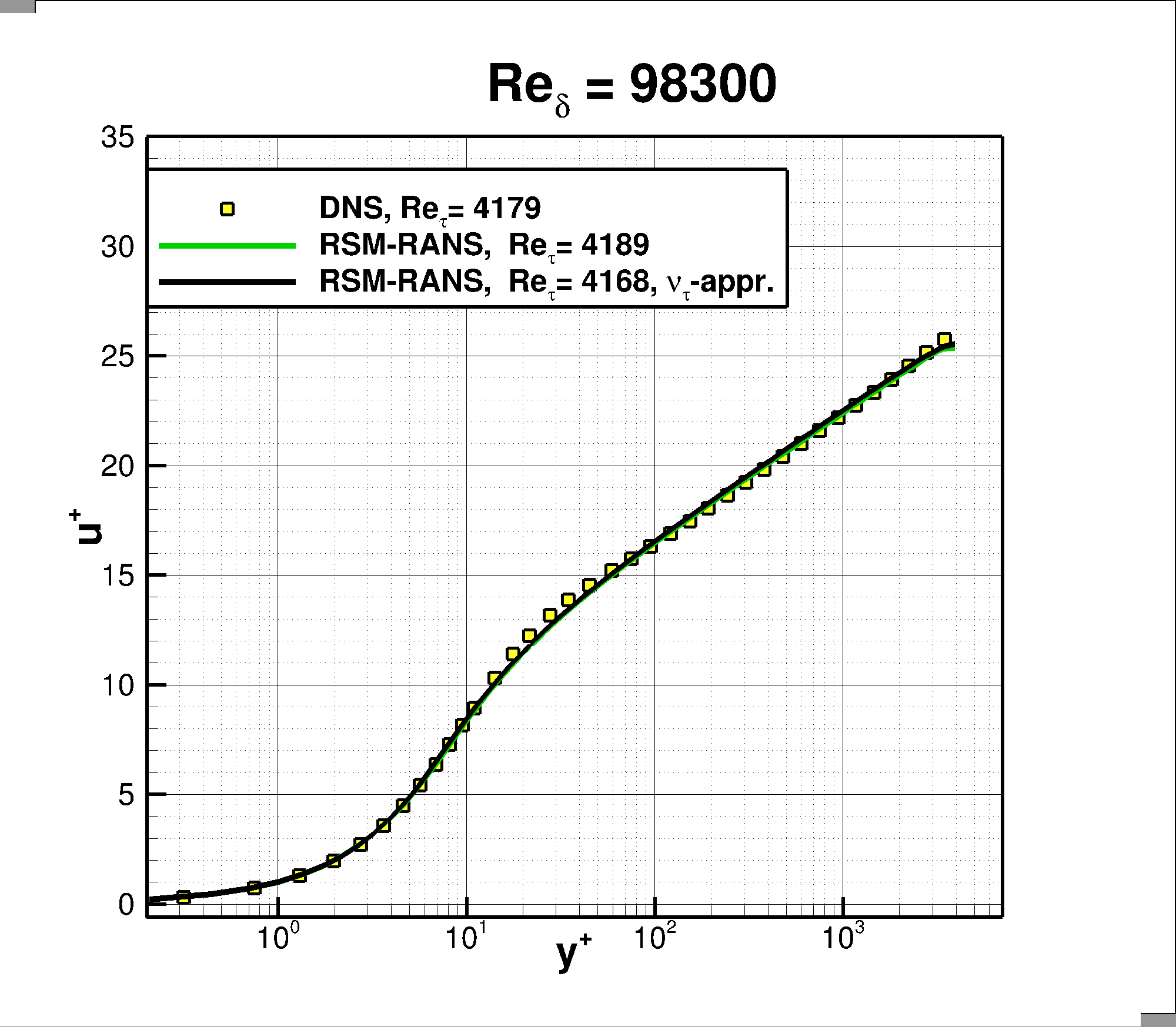}	
            \caption{RANS solutions of original SSG/LRR-$\omega$ RSM and eddy viscosity based SSG/LRR-$\omega$ RSM.}
            \label{abb:periodic_channel_rans_vergleich}
\end{subfigure}  \hspace*{0.3cm}
 \begin{subfigure}[c]{0.48\textwidth}
			\includegraphics[trim= 110 90 100 80, clip,width=\linewidth]{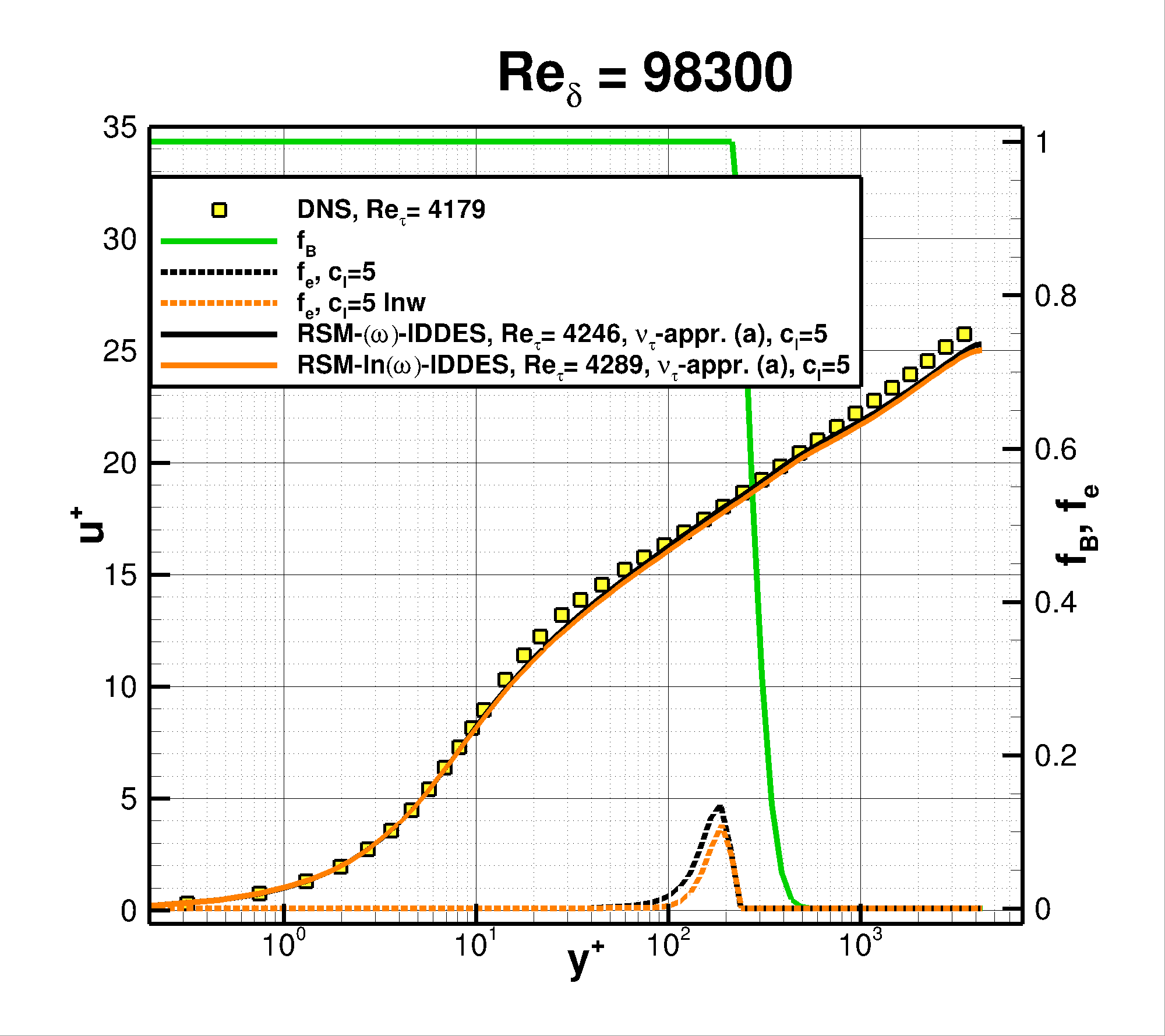}	
	        \caption{Comparison of RSM-IDDES solutions with $\omega$ as well as $\ln(\omega)$ length scale variable formulation.}
         \label{abb:periodic_channel_model_a_re_4200}
 \end{subfigure}%
\caption{Velocity profiles of periodic channel flow of RANS and RSM-IDDES solutions with a DNS reference \cite{lozano2014effect}. For the RSM-IDDES results (Fig. \ref{abb:periodic_channel_model_a_re_4200}), the corresponding IDDES model functions $f_B$ and $f_e$ are presented. The function $f_B$ indicates the transition from RANS to LES mode.}
\end{center}
\end{figure}

Prior to the RSM IDDES simulation data, results of the original RSM RANS model as well as the eddy viscosity based RSM RANS model are presented for the higher Reynolds number $Re_\delta = 98300$. Note that for the latter model, the Reynolds stress tensor $\tau^F_{i,j, RANS}$ of the RANS momentum equations is expressed by the Boussinesq approximation (Eq. \ref{eq:boussinesq}) in the entire flow domain. 
As visible in Fig. \ref{abb:periodic_channel_rans_vergleich} both RANS solutions agree very well with the DNS reference velocity profile as well as the DNS friction Reynolds number $Re_{\tau}$. 
This result demonstrates that the novel sub-grid model for IDDES is in principle also suitable for pure RANS modelling.

\subsubsection{Results for model variant (a)}
\label{sec:channel_variant_a}
\begin{figure}
\begin{center}
\begin{subfigure}[c]{0.48\textwidth}
			\includegraphics[trim= 110 90 100 80, clip,width=\linewidth]{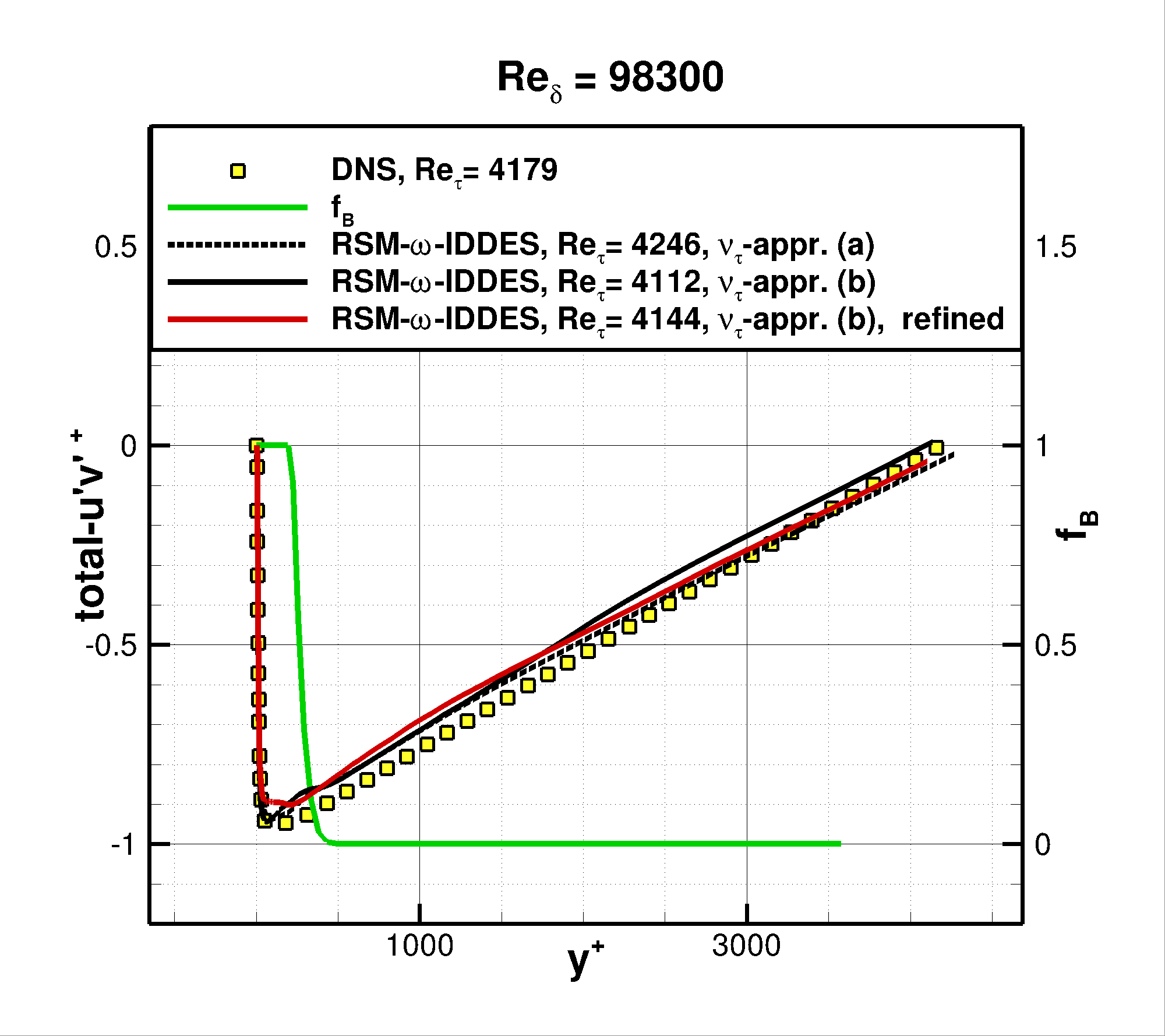}	
            \caption{}
            \label{abb:periodic_channel_model_b_re_4200_shear_stress}
\end{subfigure}  \hspace*{0.3cm}
 \begin{subfigure}[c]{0.48\textwidth}
			\includegraphics[trim= 70 50 70 50, clip,width=\linewidth]{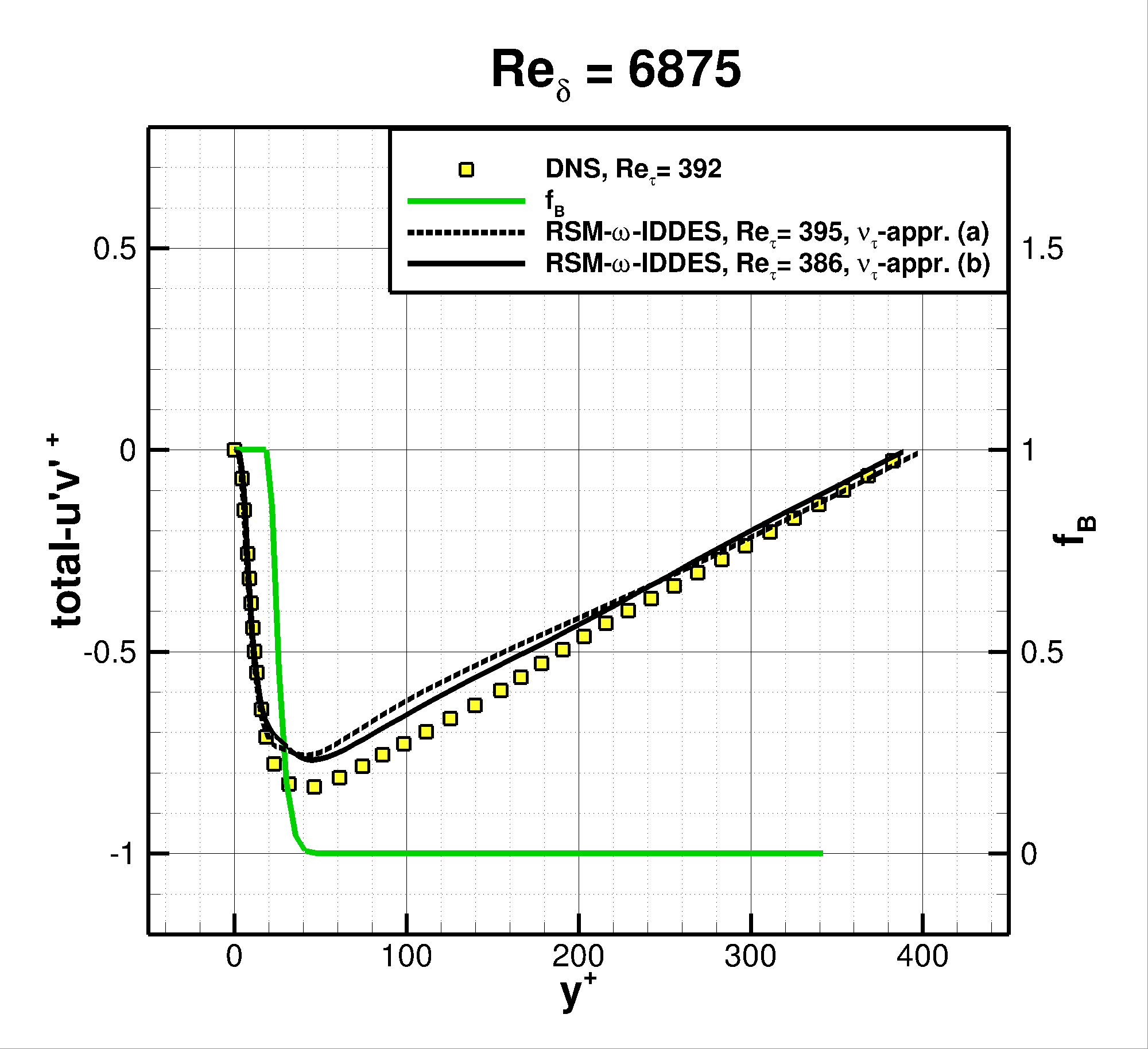}	
	        \caption{}
            \label{abb:periodic_channel_model_b_re_395_shear_stress}
\end{subfigure}%
\caption{Normalised shear stress profiles total-$u'v'^+$ of both RSM-IDDES model variants at two different bulk Reynolds numbers. The DNS reference data was obtained from \cite{lozano2014effect} and \cite{moser1999direct}.}
\end{center}
\end{figure}

Using model variant (a) as described in Sec. \ref{sec:ev_hyb_approach}, the differential SSG-LRR RSM is applied in the wall-adjacent RANS zone whereas the eddy-viscosity-based sub-grid model is employed in the LES area. As in the original IDDES formulation the spatial transition between RANS and LES zones is accomplished by the IDDES model function $f_B$ (cf. Eq.~(\ref{eq:lhyb})). 
The turbulence model specific calibrations constants $c_l$ and $c_t$ of the WMLES branch of IDDES were initially adopted from the SST-IDDES method and thus defined as $c_l=5$ and $c_t=1.87$.
\paragraph{Results for $Re_\delta = 98300$}
In the following, model variant (a) is investigated for a constant bulk Reynolds number of $Re_\delta = 98300$ and the resulting friction Reynolds numbers $ Re_{\tau}$ as well as velocity and shear stress profiles are analysed.
Initially, results of the temporal and spatial averaged friction Reynolds numbers $ Re_{\tau}=U_{\tau}\cdot\delta/\nu $, which is based on the friction velocity $U_{\tau}$, are considered. Compared to $Re_{\tau}=4179$ of the DNS reference, the friction Reynolds numbers amount to $Re_{\tau}=4246$ for RSM-$\omega$-IDDES and $Re_{\tau}=4289$ for RSM-$\ln(\omega)$-IDDES, and hence, are slightly overestimated.
Fig. \ref{abb:periodic_channel_model_a_re_4200} shows the normalised velocities $u^+=u/U_{\tau}$ over the normalised wall distance $y^+=U_{\tau}d_w /\nu$, where $U_{\tau}$ was calculated from the respective WMLES results. 
The shapes of both velocity profiles agree overall well with the reference DNS. The underestimation of the normalised velocities is due to the normalisation of $u^+$ with $U_{\tau}$ and thus a result of the overestimation of $Re_{\tau}$.
Furthermore, a corresponding total shear stress profile total-$u'v'^+$, which is defined as the sum of modelled shear stress $\widetilde{u''v''}^+$ (cf. Eq. \ref{eq:boussinesq}) and resolved shear stress $\overline{u'v'}^+$, is presented in Fig. \ref{abb:periodic_channel_model_b_re_4200_shear_stress}. The shear stresses are normalised with the squared friction velocity $U_\tau$.
The curve of total-$u'v'^+$ shows fairly good agreements with the DNS reference.

\paragraph{Calibration of IDDES model constants $c_l$ and $c_t$}
The WMLES branch of the original IDDES formulation \cite{shur2008hybrid} contains two calibration constants $c_l$ and $c_t$, which are dependent on the underlying RANS sub-grid model.
These constants are included in the elevating function $f_e$, which is constructed in order to reduce a damping of the modelled Reynolds stresses in the RANS-LES intersection region (cf. \ref{app:iddes} for more details).
To assess the impact of the $c_l$ and $c_t$ calibration constants on the $f_e$ function as well as the actual flow results, sensitivity studies have been performed. 
Starting point of these investigation has been the parameter set $\{c_l=5;  c_t=1.87\}$.

\begin{figure}
\begin{center}
\begin{subfigure}[c]{0.48\textwidth}
			\includegraphics[trim= 110 90 100 80, clip,width=\linewidth]{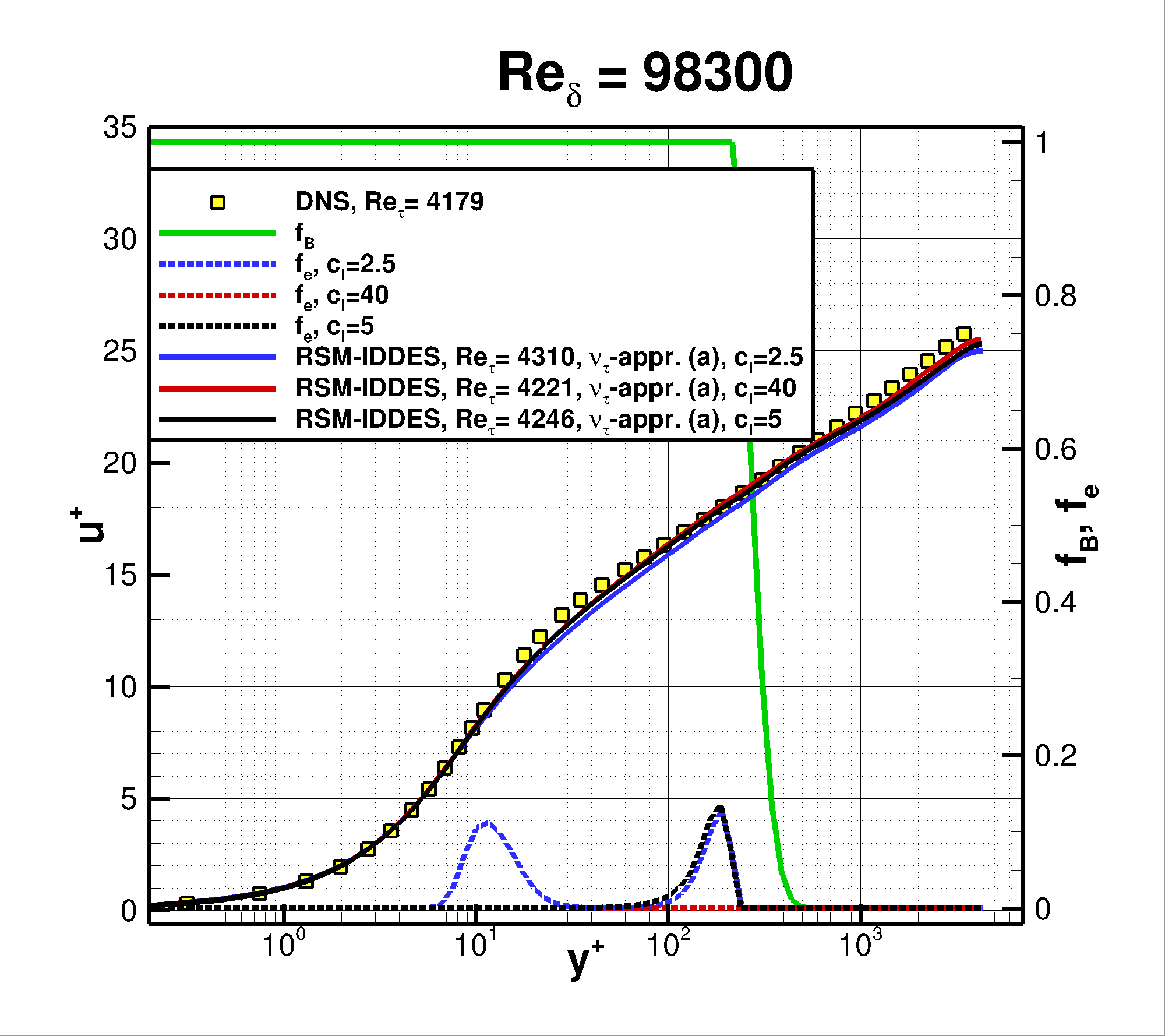}	
            \caption{}
\end{subfigure}  \hspace*{0.3cm}
 \begin{subfigure}[c]{0.48\textwidth}
			\includegraphics[trim= 110 90 100 80, clip,width=\linewidth]{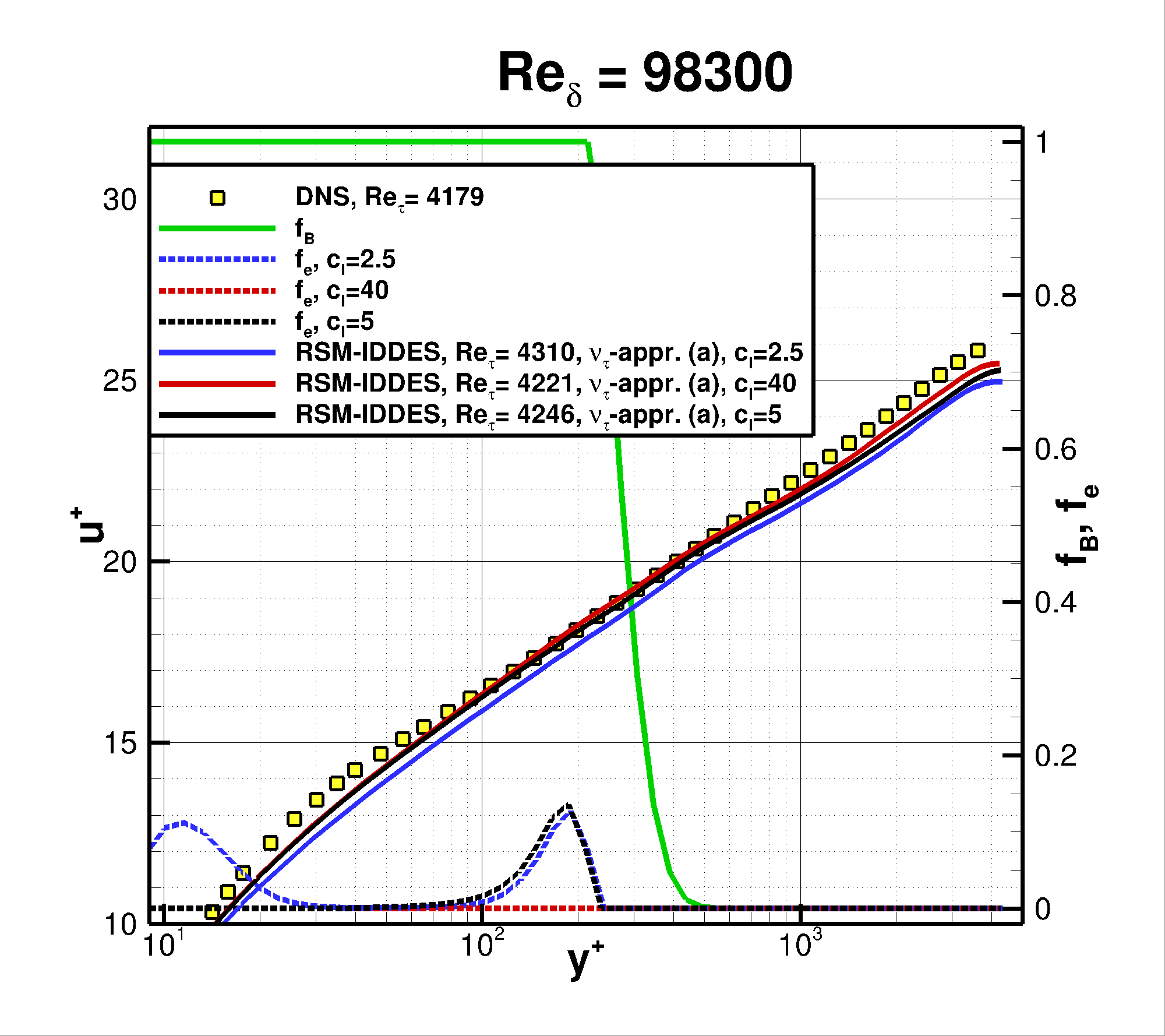}	
	        \caption{Close up view of left Fig.}
 \end{subfigure}%
\caption{Velocity profiles $u^+$ and IDDES model functions $f_B$ and $f_e$ obtained by model variant (a) in conjunction with the DNS reference \cite{lozano2014effect}.}
\label{abb:periodic_channel_model_a_calibration}
\end{center}
\end{figure}

Figure \ref{abb:periodic_channel_model_a_calibration} presents simulation results 
for different values of $c_l$ ($c_t=1.87$) in addition with the corresponding model functions $f_e$.
For $c_l=5$ a single peak of $f_e$ is present around $y^+=200$ (black dashed curve) which aims to elevate the integral RANS length scale ($l_{RANS}$) in the RANS-LES intersection region (cf. Eq. \ref{eq:lhyb}).
An increase of $c_l$ to $c_l=40$ leads to a switch-off of the $f_e$ function (red dashed curve). However, this switch-off has only a minor impact on $Re_{\tau}$ as well as the velocity profile but shows comparable results as for $c_l=5$. 
This is in accordance with the results from \cite{gritskevich2012development}, where it was demonstrated that the impact of the $f_e$ function within a related SST-IDDES is of minor importance.
For a reduced value of $c_l$ ($c_l=2.5$) a second peak in $f_e$ occurs around $y^+=10$ while the peak around $y^+=200$ remains unchanged. This second peak adversely affects $Re_\tau$ (overestimation of 3.5\,\% compared to the DNS reference) and thus leads to a systematic underestimation of the velocity profile.
The presence of this second peak in  $f_e$, well outside of the RANS-LES intersection range, indicates that $f_e$ is used outside of its design point.
Simulation results of a sensitivity study of the $c_t$ calibration constant (for $c_l=5$) do not show significant changes in $f_e$. Consequently, no impact on $Re_\tau$ as well as the velocity profiles have been observed (not presented here).%

As a result, it has been shown than an adaption of the initial calibration constant $c_l=5$ and $c_t=1.87$ (adopted from SST-IDDES) does not lead to noticeably improved agreements with the DNS reference. A further suitability of these constants is presented in the following sections for a refined mesh as well as for a reduced Reynolds number of the channel flow.

\begin{figure}
\begin{center}
\begin{subfigure}[c]{0.48\textwidth}
			\includegraphics[trim= 70 50 70 50, clip,width=\linewidth]{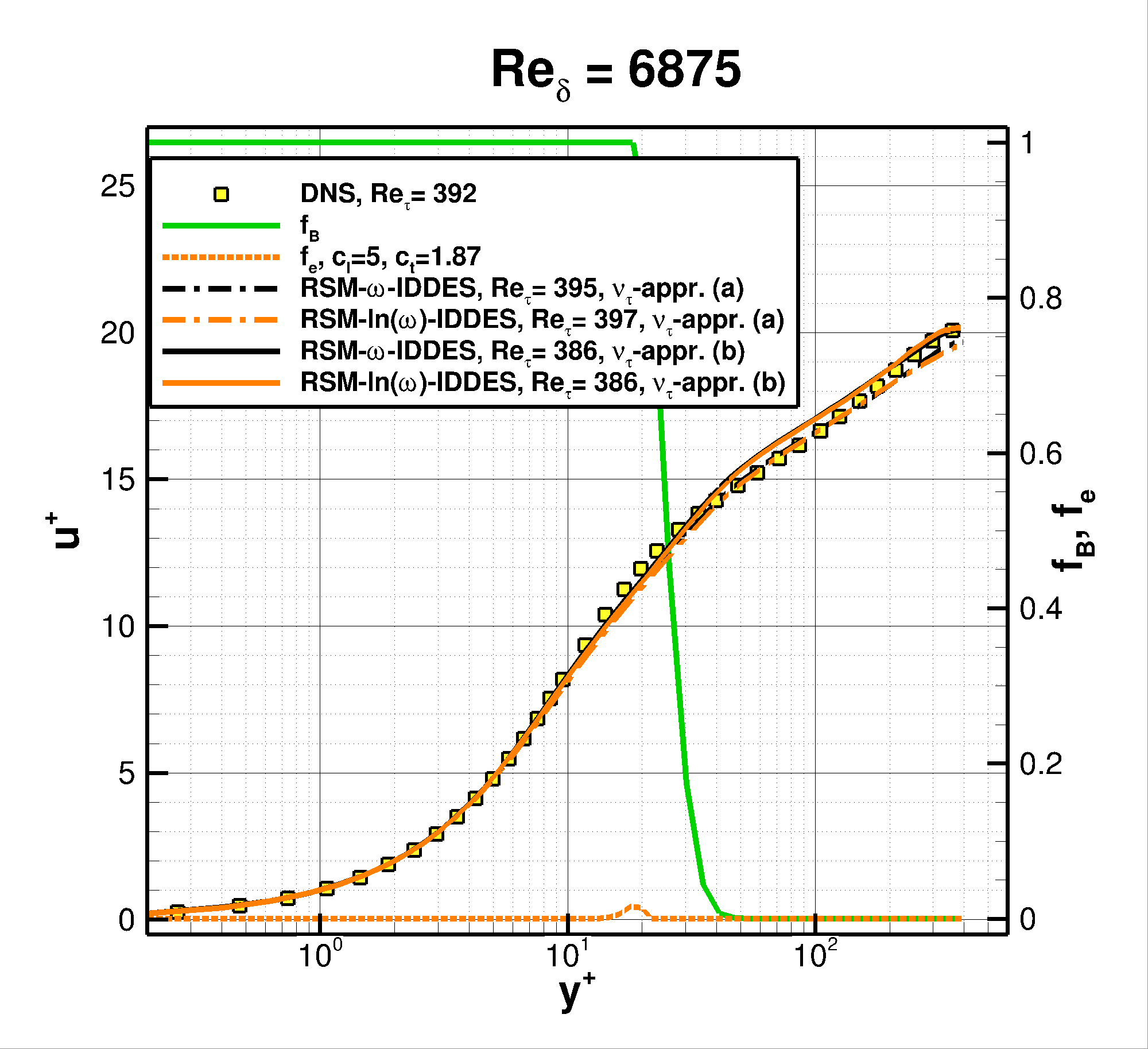}	
            \caption{}         %
\end{subfigure}  \hspace*{0.3cm}
 \begin{subfigure}[c]{0.48\textwidth}
			\includegraphics[trim= 70 50 70 50, clip,width=\linewidth]{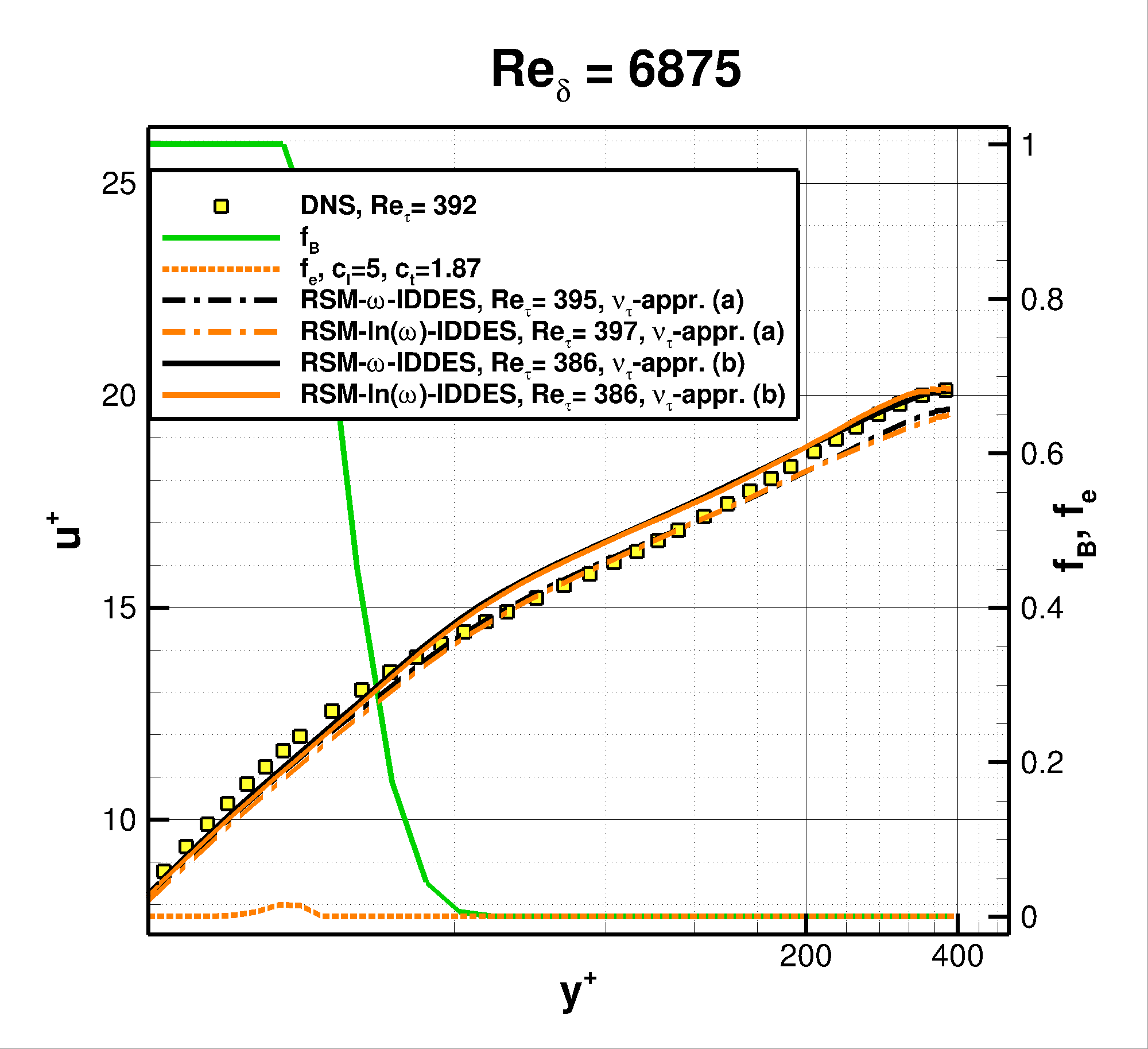}	
	        \caption{Close up view of left Fig.}
\end{subfigure}%
\caption{Velocity profiles $y^+$ for both RSM-IDDES model variants and IDDES model functions $f_B$ and $f_e$ at the lower bulk Reynolds number
with DNS reference \cite{moser1999direct}.}
\label{abb:periodic_channel_model_b_re_395_velocity}
\end{center}
\end{figure}

\paragraph{Results for $Re_{\delta}=6875$}
Simulations at a much (14 times) smaller Reynolds number of $Re_{\delta}=6875$ while using the same wall-tangential mesh resolution $\Delta x$ and $ \Delta z$ leads to an effective refinement of $\Delta x ^+ $ and $ \Delta z^+$ close to wall-resolved LES requirements (cf. \cite{Probst2016}).
This in turn causes an earlier transition of the $f_B$ function from RANS to LES mode in terms of $y^+$. The transition already starts at $y^+=20$ compared to $y^+=200$  and thus a larger fraction of the boundary layer is resolved for the smaller Reynolds number (cf. Fig. \ref{abb:periodic_channel_model_b_re_395_velocity} and \ref{abb:periodic_channel_model_b_overview_re_4200_velocity}).
As shown in Fig. \ref{abb:periodic_channel_model_b_re_395_velocity}, the values of $Re_\tau$ agree almost perfectly (overestimation of max $1.3\,\%$) with the DNS reference for both length-scale variable formulations. Furthermore, the velocity profiles agree fairly well with the DNS reference and only in the middle of the channel ($y^+ \geq 200$) minor underestimations occur, which is consistent with the results at the higher Reynolds number. 
As demonstrated in Fig. \ref{abb:periodic_channel_model_a_calibration}, turning off the $f_e$ model function results into an elevation of the velocity profile and could thus improve the agreements with the DNS reference in the present case. However, this function is already close to zero at the selected Reynolds number, so that no visible effects are to be expected (cf. Fig. \ref{abb:periodic_channel_model_b_re_395_velocity}). This confirms that the previous calibration of the constants $c_l$ and $c_t$, which control the $f_e$ function, is also suitable for small Reynolds numbers.%
Additionally, the corresponding shear stress profile total-$u'v'^+$ is presented in Fig. \ref{abb:periodic_channel_model_b_re_395_shear_stress}, which shows acceptable agreements with the DNS reference.

\paragraph{Intermediate summary}
A basic functionality of the WMLES branch with a RSM sub-grid model in conjunction with the SSG/LRR RANS model has been demonstrated. 
However, slight deviations like the
overestimated values of $Re_\tau$
motivate further investigations on the hybrid RANS-LES coupling which are presented in the following section.

\subsubsection{Results for model variant (b)}
\label{sec:channel_variant_b}

In variant (b) of the present RSM-IDDES (cf. Sec. \ref{sec:ev_hyb_approach}), the eddy-viscosity approach for the sub-grid model is also applied to the near-wall RANS layer of the WMLES branch. Thus, for the channel flow the Reynolds stress tensor in Eq.~(\ref{eq:momentum}) is replaced by the Boussinesq approximation in the entire domain. 

\begin{figure}
\begin{center}
\begin{subfigure}[c]{0.48\textwidth}
			\includegraphics[trim= 110 90 100 80, clip,width=\linewidth]{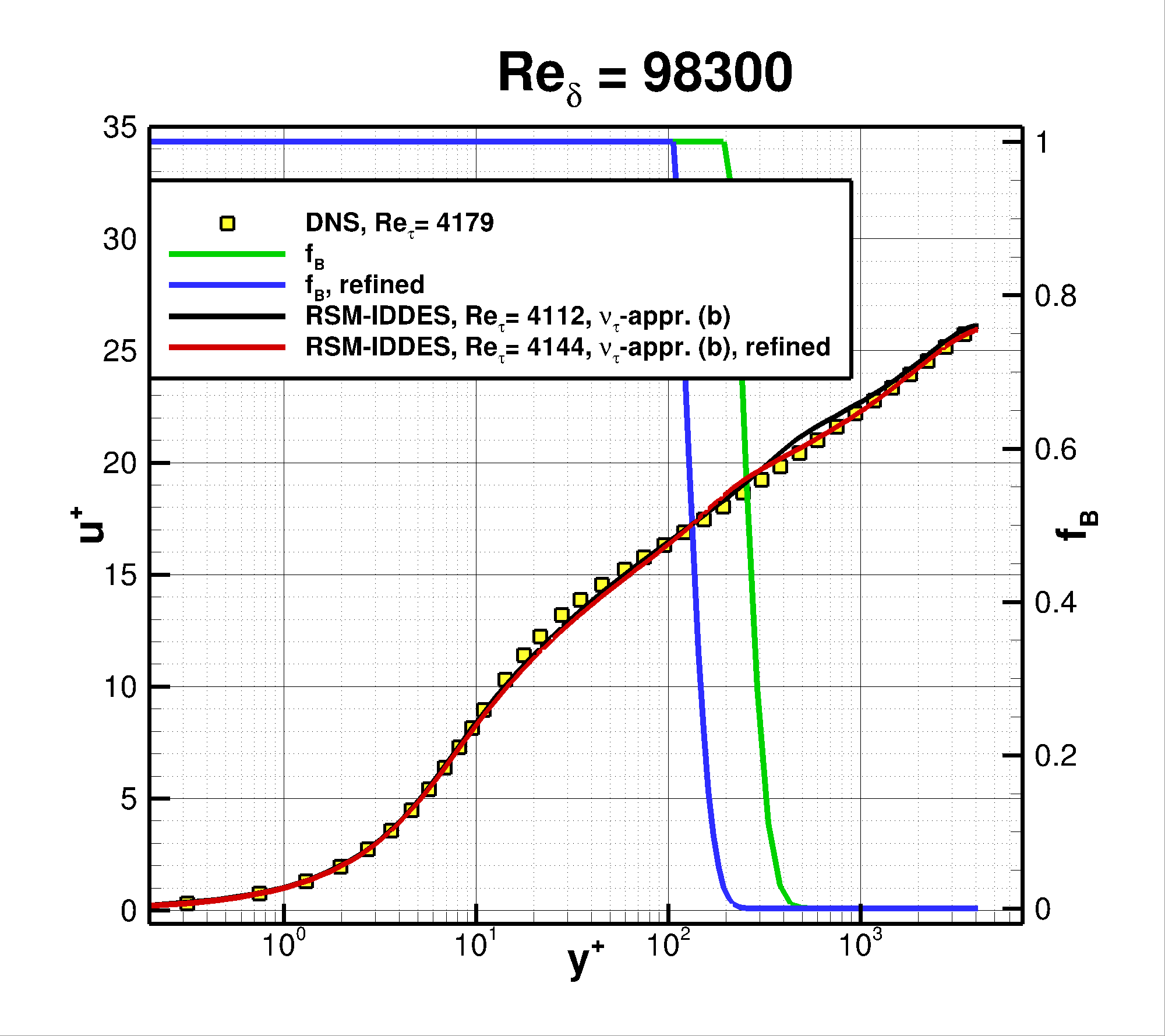}	
            \caption{}           %
\end{subfigure}  \hspace*{0.3cm}
 \begin{subfigure}[c]{0.48\textwidth}
			\includegraphics[trim= 70 50 70 50, clip,width=\linewidth]{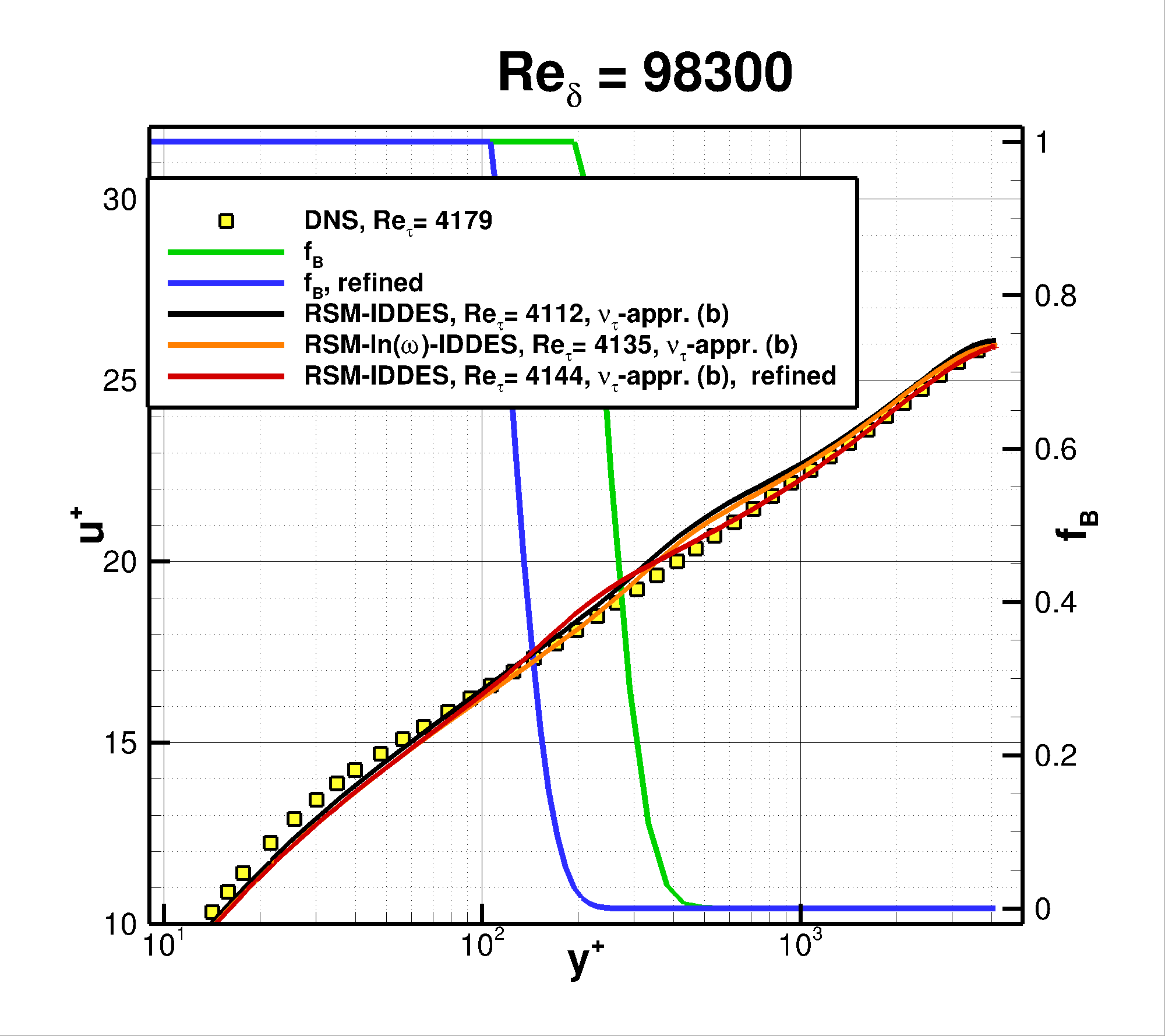}	
	        \caption{Close up view of left Fig.}
\end{subfigure}%
\caption{
Velocity profiles $y^+$ for model variant (b).}
\label{abb:periodic_channel_model_b_overview_re_4200_velocity}
\end{center}
\end{figure}

\paragraph{Results for $Re_{\delta}=98300$}
In the following, results of the RSM-IDDES for both length-scale variable formulations $\omega$ and $\ln (\omega) $ are presented.
The corresponding friction Reynolds numbers $Re_{\tau}=4112$ ($\omega$) and $Re_{\tau}=4135$ ($\ln (\omega))$ are both just slightly underestimated  by $-1.5\,\%$ and $-1.0\,\%$ compared to DNS data. Furthermore, the shape and absolute values of the corresponding velocity profiles in Fig. \ref{abb:periodic_channel_model_b_overview_re_4200_velocity}, show very good agreement with the DNS reference (black and orange curves). %

In order to identify a potential mesh dependence of the flow solution a global mesh refinement study was performed. To this end, the mesh resolution was increased by a factor of two in all spatial coordinate directions resulting in 130 x 282 x 130 cells in $x$, $y$ and $z$, respectively. The averaged friction Reynolds number amounts $Re_{\tau}=4144$ which represents an underestimation of only $-0.7\,\%$ compared to the DNS reference. In addition to that, the velocity profile is very close to the DNS data in the entire LES regime (red curve in Fig. \ref{abb:periodic_channel_model_b_overview_re_4200_velocity}). 
Additionally, normalised shear stress profiles total-$u'v'^+$ of the previous simulations are depicted in Fig. \ref{abb:periodic_channel_model_b_re_4200_shear_stress}. All curves shows good agreements with the DNS reference.

\paragraph{Results for $Re_{\delta}=6875$}

The numerical results using model variant (b) of the present RSM-IDDES show similar trends as for the higher Reynolds number. 
Again, good agreement between RSM-$\omega$-IDDES and RSM-$\ln (\omega)$-IDDES is obtained with respect to $Re_{\tau}$ and the velocity profiles (cf. Fig \ref{abb:periodic_channel_model_b_re_395_velocity}).
Compared to the DNS reference, the $Re_{\tau}$ values are slightly underestimated by $-1.8\,\%$ for both length-scale variable formulations. 
Furthermore, the corresponding shear stress profile total-$u'v'^+$ is depicted in Fig. \ref{abb:periodic_channel_model_b_re_395_shear_stress} showing acceptable agreements with a DNS reference.

\paragraph{Intermediate summary}
The overall good agreements of model version (b) with the DNS reference validates the previous calibration of the IDDES model coefficients $c_t$ and $c_l$. Thus, the initial values of $c_t=1.87$ and $c_l=5$ originating from the SST-IDDES model are used in the novel RSM-IDDES model.

\subsection{Embedded WMLES functionality}\label{sec:flat_plate}
In this section the flow on a flat-plate boundary layer with zero pressure gradient is considered. In contrast to the periodic channel, the (mean) flow quantities also depend on the streamwise coordinate $x$. 
With regard to the computational setup the hybrid RANS-LES methodology is locally embedded in a surrounding RSM-RANS region, referred to as embedded WMLES (cf. Sec.~\ref{sec:stg}). 
Similar to the periodic channel flow, we did not employ the automatisms of the IDDES (automatic switch to RANS, DDES or WMLES modes). 
Instead, the integral length scale $l_{\text{RANS}}$ of the RANS turbulence model (cf. Eq. \ref{eq:lrans}) remained unchanged in selected flow regions and was not replaced by a $l_{DES}$ length scale. Thus, RANS regions outside the mesh refinement area were defined manually.
Additionally, the same zonal definition is applied to the stress blending function 
$\Tilde{f}_{user}$ in Eq.~(\ref{eq:blending_rst}) in order to realise variant (b) of RSM-IDDES in the embedded region.

The boundary layer starts with zero thickness at the inflow of the computational domain and is modelled using RSM-RANS (SSG/LRR) until the momentum-thickness Reynolds number reaches $Re_{\theta} = 3040$. 
From this point, $l_{\text{RANS}}$ from Eq. \ref{eq:lrans} is replaced by the IDDES length scale $l_{\text{hyb}}$ (manual switch from RANS to IDDES).
As for the periodic channel, $f_{dt}$ is set to 1 in order to enforce the WMLES branch of IDDES. Thus $l_{\text{hyb}}$ simplifies to $l_{\text{WMLES}}$ (cf. \ref{app:iddes}).
A synthetic-turbulence forcing region is applied where $Re_{\theta} = 3040$ to achieve a rapid streamwise transition from RANS to WMLES. The synthetic turbulence generator (STG) is active in a streamwise domain of about half of local boundary layer thickness $\delta_{STG}$.
In spanwise direction the spatial extent amounts to 5$\,\delta_{STG}$, and periodic boundary conditions are applied.

With respect to the discretization strategy, a hybrid grid with structured hexahedral cells within the WMLES area and partly unstructured cells in the surrounding RANS area is employed.
The WMLES part of the grid is derived from a fully-structured grid for the same flow conditions used in \cite{Deck2018}.
The resolution within this area is designed to fulfil
$\delta_{x} / \Delta x =10 $ as well as $\delta_{x} / \Delta z= 20 $
throughout the entire WMLES domain, where $\delta_x$ represents the local boundary thickness. 
The wall-normal grid spacing of wall adjacent cells is limited to $\Delta y ^+ = 0.2$ and safely fulfils the demands of the RSM RANS model. 
With these resolution criteria, the entire grid comprises 8.3 million points.
The normalised time step is set to $\Delta t^+=0.4$ and satisfies the convective CFL criterion CFL$\leq 1$ throughout the entire LES regime. The developing boundary layer was simulated about $2.5$\,CTU (with respect to length of the WMLES domain) where $1.2$\,CTU are used for the calculation of time-averaged quantities along with additional spanwise averaging.

\paragraph{Results of skin friction distribution mean-$c_f$}
The simulation results of the evolving flat-plate boundary layer are presented as surface distributions of the time- and span-averaged skin friction coefficient mean-$c_f$. 

\begin{figure}
\begin{center}
\begin{subfigure}[c]{0.48\textwidth}
			\includegraphics[trim= 20 40 150 200, clip,width=\linewidth]{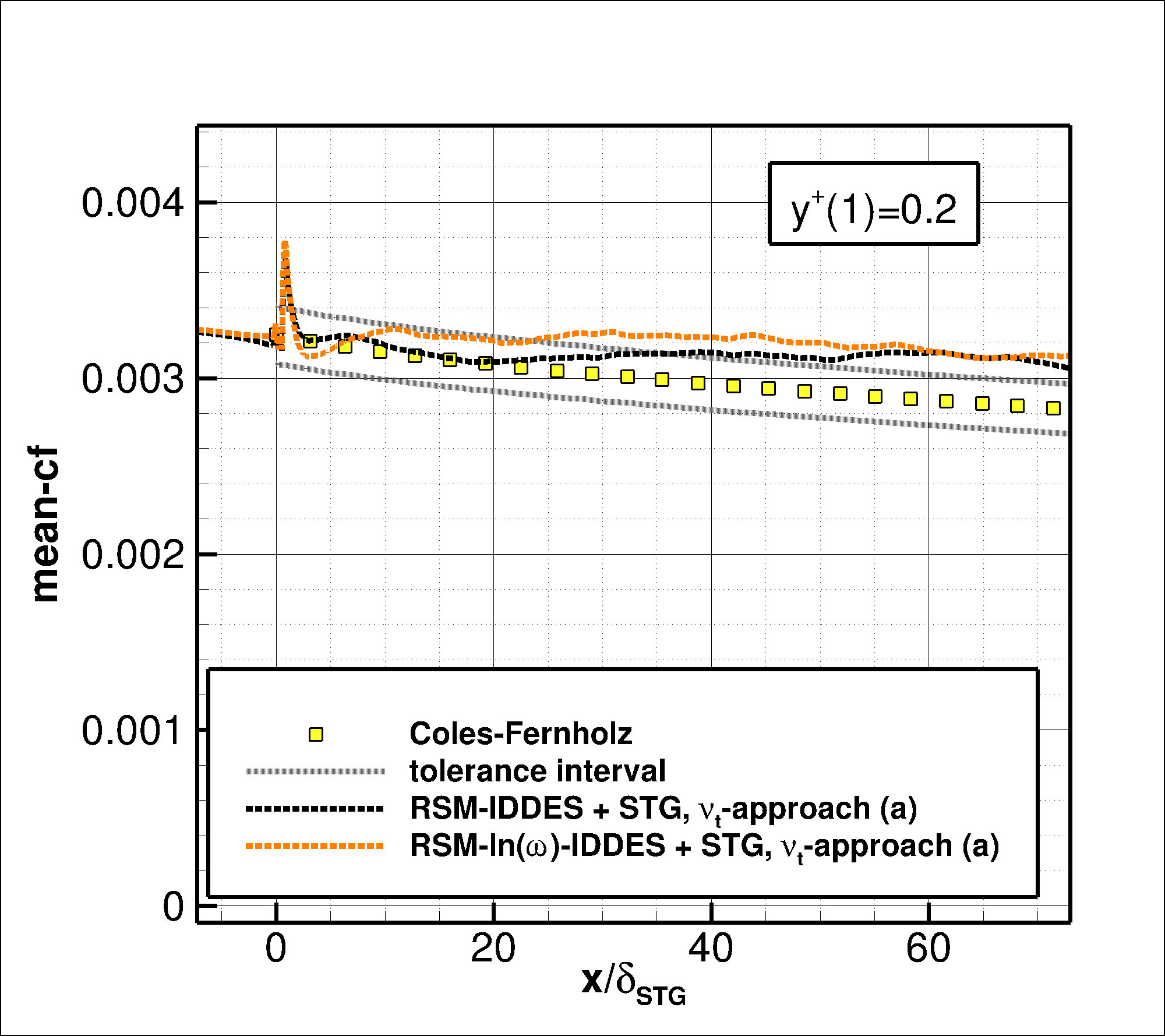}	
            \caption{Results of RSM-IDDES $\nu_t$-approach (a)}
            \label{abb:flat_plate_model_a}
            \vspace{0.32cm}
\end{subfigure}  \hspace*{0.3cm}
 \begin{subfigure}[c]{0.48\textwidth}
			\includegraphics[trim= 20 40 150 200, clip,width=\linewidth]{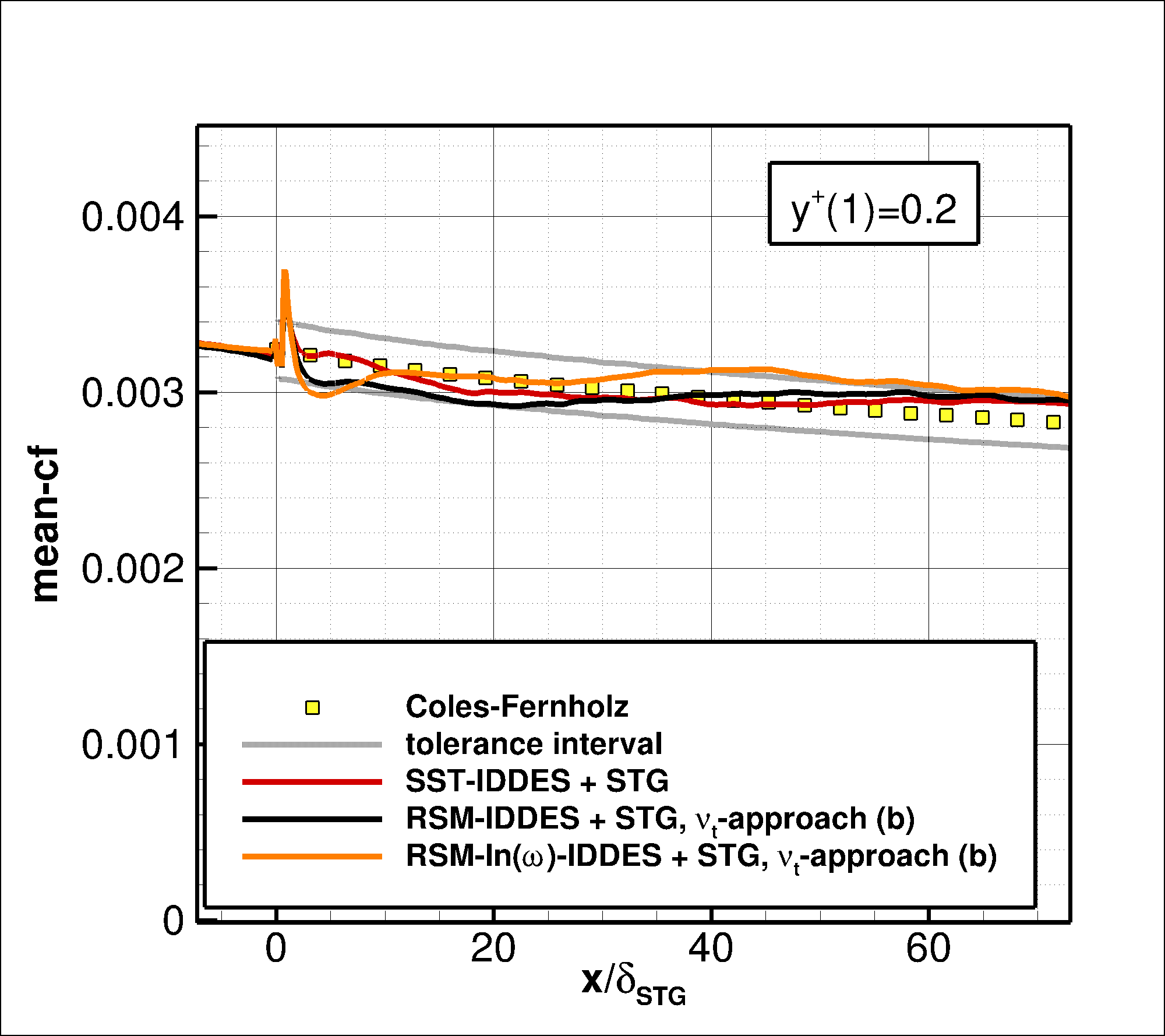}	
	        \caption{Results of RSM-IDDES $\nu_t$-approach (b) with SST-IDDES reference data.}
            \label{abb:flat_plate_model_b}
\end{subfigure}%
\caption{Skin-friction distributions of the flow about a flat plate with Coles-Fernholz correlation \cite{nagib2007approach}.}
\end{center}
\end{figure}

Results of the RSM-IDDES of model version (a) are shown in Fig.~\ref{abb:flat_plate_model_a} along with the Coles-Fernholz correlation \cite{nagib2007approach}.
It is noticeable that for both formulations of the length scale variable, the mean-$c_f$ curves leave the $5\,\%$ tolerance interval (dashed curves) as $x$ increases, leading to a clear overestimation of mean-$c_f$.
However, a significant overestimation of mean-$c_f$ for the fully developed turbulent boundary layer ($x/\delta_{STG} \geq 60$) is also expected for model version (a). This is due to the already observed overestimation of $Re_\tau$ for the periodic channel flow (cf. Fig. \ref{abb:periodic_channel_model_a_re_4200}). Since mean-$c_f$ behaves proportional to $Re_{\tau}^2$, deviations in $Re_{\tau}$ are more pronounced in the mean-$c_f$ plot.
In contrast to the expected decrease of the skin friction in streamwise direction (as present for the Coles-Fernholz correlation), its level remains at an almost constant value $c_f \approx 0.032$ after an typical initial overshoot. The latter is due to the injected turbulence from STG at the RANS-LES interface \cite{Probst2020}. 
With regard to the comparison of RSM-$\omega$-IDDES and RSM-$\ln(\omega)$-IDDES, visible differences appear downstream of the STG but vanish with increasing $x$ and almost align at the end of the flat plate. This behaviour is also present for model version (b) (cf. Fig. \ref{abb:flat_plate_model_b}).
Given the very good agreement between the two formulations of the length scale variable in the previous test cases, these differences can probably be attributed to interactions between the STG and the respective length scale formulation. 
This would be consistent with subsequent results of sensitivity studies, where varying injections of resolved turbulence only have a particular effect on the flow region directly downstream. For larger distances, however, negligible differences are present.

The results of the RSM-IDDES with model version (b) are presented in Fig. \ref{abb:flat_plate_model_b} with SST-IDDES data of the same numerical setup. Despite visible deviations of the RSM-IDDES compared to the Coles-Fernholz correlation, the skin friction distribution mostly remains within the $5\,\%$ tolerance interval and therefore indicates acceptable results. 
As for model version (a) the mean-$c_f$ level remains at an almost constant value.
However, the comparison results from SST-IDDES reveal very similar behaviour with an nearly constant mean-$c_f$ level at 0.003.
Thus, it can be stated that the present RSM-IDDES behaves widely consistent with a well-established reference model for this particular flow and numerical setup.
Note that SST-based IDDES was chosen as reference because of many conceptual similarities between the SST and SSG/LRR-RSM RANS models, i.e. both sharing the same $\omega$ length-scale equation and blending functions to switch between different modelling regimes \cite{Eisfeld2016}.
In contrast, earlier investigations of this flow using Spalart-Allmaras-based IDDES and STG show a better agreement with the expected mean-$c_f$ decrease \cite{Probst2020}.
Therefore, a more fundamental root cause of the deviations, probably related to $\omega$-based IDDES modelling or its interaction with STG, is presumed, rather than specific issues with the present coupling to RSM.

\paragraph{Sensitivity study on RANS-LES transition}
The unexpected behaviour of the skin friction development motivates 
targeted sensitivity studies on the modelling of the RANS-LES transition related to the STG injection at $x_{\delta_{STG}}= 0$.
Note, that the following investigations were performed on a mesh with a coarsened $y^+$ value of 1 in order to reduce the simulation time. 
This slightly affects the absolute level of mean-$c_f$, but not the qualitative development and can therefore be accepted for this study.

\begin{figure}
\begin{center}
\begin{subfigure}[c]{0.48\textwidth}
			\includegraphics[trim= 20 40 150 200, clip,width=\linewidth]{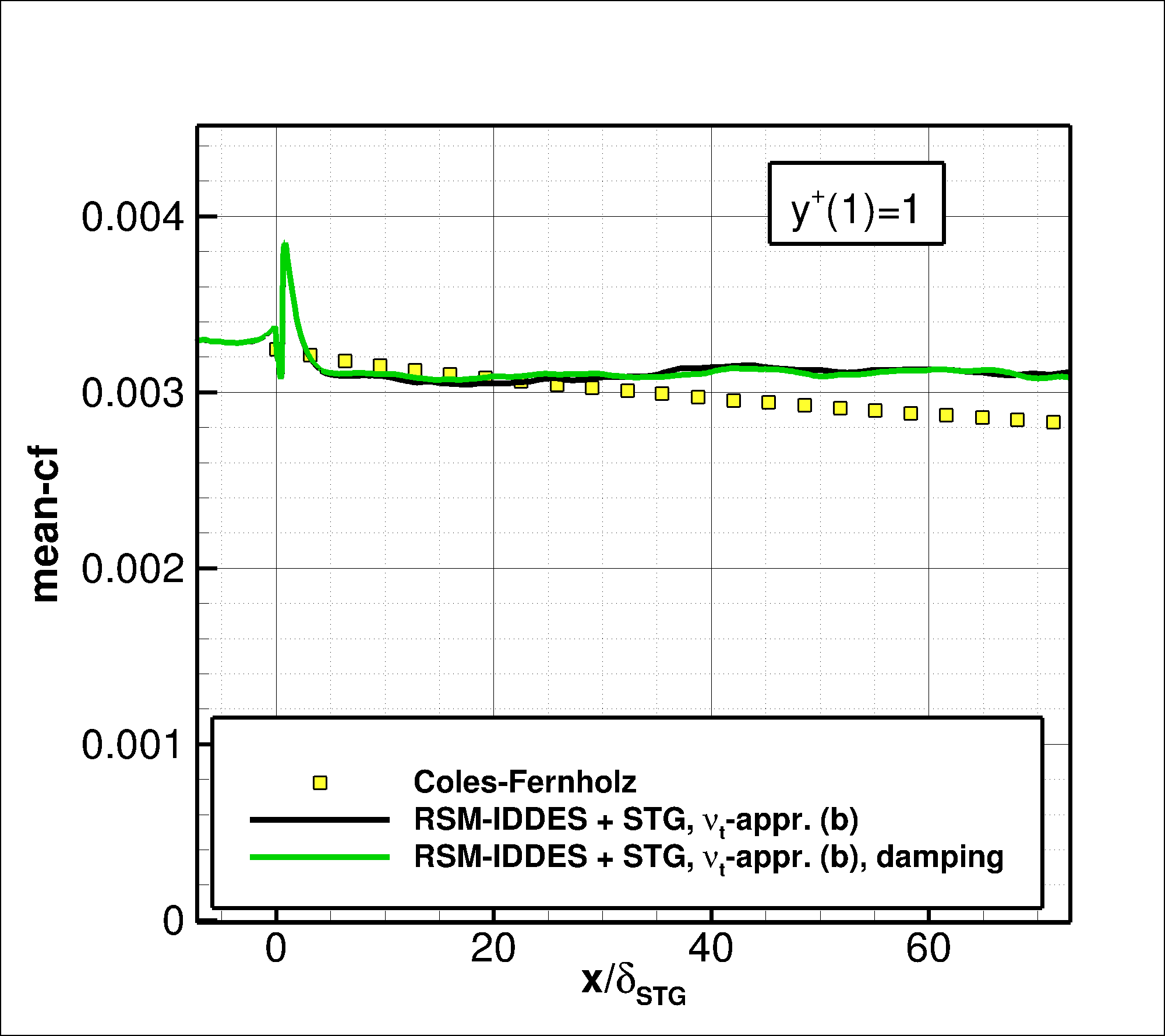}	
	        \caption{Impact of effective damping of convected RANS turbulence.}
            \label{abb:flat_plate_sink_term}
\end{subfigure} \hspace*{0.3cm}
\begin{subfigure}[c]{0.48\textwidth}
			\includegraphics[trim= 20 40 150 200, clip,width=\linewidth]{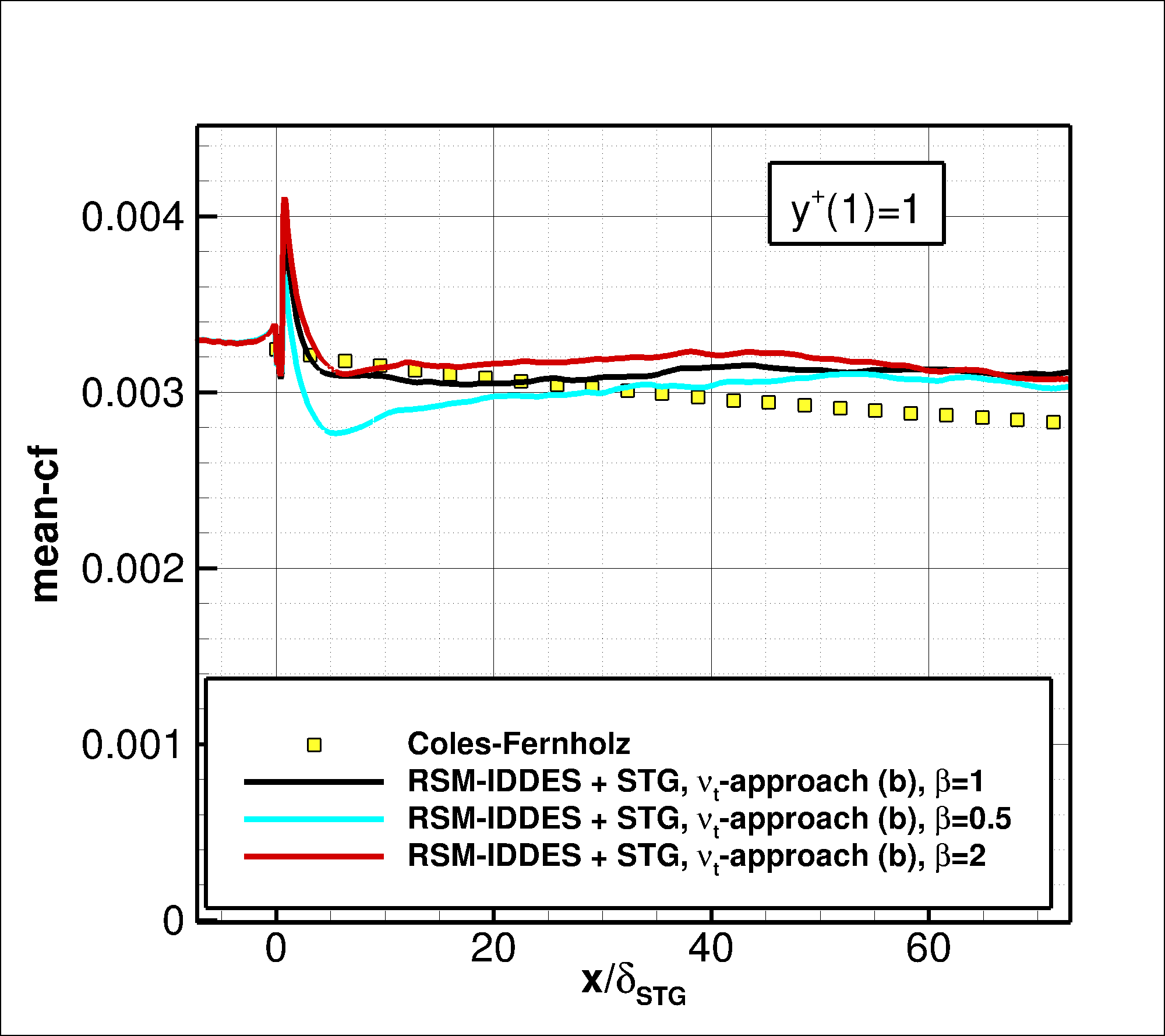}	
	        \caption{Impact of varied scaling parameter for the STG length scale $l_e$ (cf. Eq.~(\ref{eq:STG_le}))}
           \label{abb:flat_plate_amplification_factor}
\end{subfigure}%
\caption{Skin-friction distributions of the flow about a flat plate.}
\end{center}
\end{figure}

One potential influencing factor on the evolution of mean-$c_f$ is the convection of rather large modelled turbulence from the RANS region into the WMLES area. 
To assess this effect, an additional damping term for the modelled 
normal Reynolds-stress components, inspired by the corresponding term for the turbulent kinetic energy in the SST-based volumetric version of STG (VSTG) \cite{Shur2017}, is added.
This term quickly drives the modelled stress towards the level of an explicit Smagorinsky sub-grid model downstream of the interface. 
However, the results of this approach depicted in Fig.~\ref{abb:flat_plate_sink_term} show no notable change compared to the basic RSM-IDDES computation. 
Thus, the convection of modelled RANS turbulence into the WMLES domain does not play a significant role on the subsequent mean-$c_f$ development in this setup.

A further investigation addresses the length scale $l_e$ used in the formulation of STG, cf. Eq.~(\ref{eq:STG_le}), which controls the size of  the synthetic structures and is related to the dissipation rate of the injected turbulence.
Compared to e.g. the Reynolds-stress input tensor used to scale the fluctuations, the length scale is considered a more uncertain modelling parameter in the synthetic-turbulence method. 
To assess its sensitivities on the results, a simple scaling factor $\beta$ for the local values of $l_e$ is introduced and varied as $\beta \in \{0.5, 1, 2\} $. Note that the variation is arbitrary and only used to reveal any effect of modifying the STG scales on the mean-$c_f$ development.

Results of the sensitivity study of $\beta$ are shown in Fig. \ref{abb:flat_plate_amplification_factor}.
A significant effect of the $\beta$ variation is visible in the flow area $0\leq x/\delta_{STG}\leq 50$. For $x/\delta_{STG}\geq 50$, however, all mean-$c_f$ curves converge and show a similar development at a constant offset from the Coles-Fernholz correlation.
With careful inspection, one may note that the slope of the mean-$c_f$ curves in this region is also rather consistent with the reference.
One possible conclusion could be, that the seemingly almost constant mean-$c_f$ value for $\beta = 1$ is actually a result of the rather wide-ranging impact of the synthetic-turbulence method which interferes with the WMLES mode of RSM-IDDES over a long distance.
However, the origin of the final offset from Coles-Fernholz, which is also present in SST-IDDES computations (cf. Fig.~\ref{abb:flat_plate_model_b}), calls for further investigation in future studies.

\section{Conclusions}\label{sec:conclusion}
A Reynolds-stress based sub-grid model for a hybrid RANS-LES method was developed. To this end, the well-known IDDES method served as foundation for the hybridisation approach. 
In the presented study, a systematic investigation of the individual sub-functionalities of IDDES was performed with the aid of suitable test cases. 

A main finding of this work was that a direct coupling of a differential RSM-RANS model with IDDES by just replacing the RANS length scale  with the provided hybrid length scale from IDDES is not suitable. It was shown that the resulting differential sub-grid LES is not capable to sufficiently dissipate the turbulent kinetic energy of the small turbulent scales in decaying isotropic turbulence (DIT) and motivated a modification of the sub-grid model.

The local replacement of the Reynolds-stress tensor of the RANS background model by the Boussinesq approximation and thus introducing an eddy viscosity as sub-grid model was shown to  provide appropriate dissipation capabilities. 
This new model was calibrated for the same DIT case and showed excellent agreement with experimental data.

Regarding the integration of the sub-grid model into the framework of IDDES, two different approaches were investigated, which differ in the formulation of the Reynolds-stress tensor in the near wall RANS layer of the wall-modelled LES branch. Both approaches showed good results in periodic channel flows at largely different bulk Reynolds numbers.
Moreover, the newly developed  RSM-IDDES was validated for two different length-scale variable formulations of the underlying SSG/LRR RSM, $\omega$ and $\ln (\omega)$, which both show consistent behaviour for all considered flow types in this study.

Although the final application in an embedded WMLES of a flat-plate boundary layer, the new model showed deviations from experimental data, the results are still widely consistent with reference data from SST-IDDES. 
With the aid of targeted sensitivity studies it was shown that these deviations can partly be traced back to the employed synthetic turbulence generator. 

Future research will focus on coupling the preferred RSM-IDDES variant (b) with IDDES in a fully-automatic non-zonal manner in order to avoid additional user input.

\section{Acknowledgments}
The authors gratefully acknowledge the Deutsche 
Forschungsgemeinschaft DFG (German Research Foundation) for funding 
this work in the framework of the research unit FOR 2895.
Additionally, the authors gratefully acknowledge the computing time granted by the Resource Allocation Board and provided on the supercomputer Lise and Emmy at NHR@ZIB and NHR@Göttingen as part of the NHR infrastructure. The calculations for this research were conducted with computing resources under the project nii00164.

\appendix
\section{IDDES method}
\label{app:iddes}

Unless otherwise stated, the governing equations of the original IDDES method \cite{shur2008hybrid} are presented below.

\paragraph{Filter width}
The IDDES method employs a filter width $\Delta $ which takes into account the wall distance $d_w$ in addition to the local grid spacings $h_{max}, h_{wn}$:
\begin{align}
\label{eqn:delta_iddes}
& \Delta = \Delta_{\mbox{\tiny IDDES}}  = \min \left\{ {\max \left[ {C_w \cdot d_w , C_w \cdot h_{\max }, h_{wn} } \right], h_{\max} } \right\}, \\
& h_{\text{max}}=\max{ \{   h_x, h_y, h_z\} }, \ \ \ C_w=0.15
\end{align}
For large values of the wall-normal grid resolution $h_{wn}$, the formulation in Eq. \ref{eqn:delta_iddes} allows to switch to an alternative function of $\Delta$ with a steeper increase of $\Delta$ in wall normal direction.

\paragraph{DDES Branch}
The DDES branch of IDDES only becomes active if the inflow conditions do not contain resolved turbulence. The formulation of the DDES length scale for RSM-IDDES was adopted from \cite{Probst2011}:

\begin{align}
    l_{\text{DDES}}&= l_{\text{RANS}} - f_d \max{\{    0, \left( l_{\text{RANS}} - l_{\text{LES}}  \right) \}} \\
 \text{with}  \ \ \   f_d &= 1-\tanh{\left[{\left(16r_d\right)}^3 \right] }  \ \ \ \\
  \text{and} \ \ \    r_d &= \frac{\nu + \nu_t}{ \kappa^2 d^2_w \cdot   {[ \sum_{i,j} {\left( \partial u_i/ \partial x_j \right)}^2 ]}^{1/2} } \ \ \ .%
\end{align}
The function $r_d$ serves as marker for different regions within the turbulent boundary layer. 
For the RSM sub-grid model, the eddy viscosity variable $\nu_t$ is calculated by the modelled turbulent kinetic energy $\Tilde{k}$ and the turbulent length scale $\omega$ provided by Menter's BSL $\omega$ equation (cf. Eq. \ref{eq:omega_transport_equation}).
\begin{align}
    \nu_t &= \frac{\Tilde{k} }{\omega} = \frac{\widetilde{v_i''v_i''} }{ 2 \omega}   \ \ .
\end{align}
As with all DES methods, $l_{RANS}$ of the RSM sub-grid model is replaces by $l_{DDES}$, where $l_{RANS}$ is defined as $l_{\text{RANS}} = {\Tilde{k}}^{1/2} / c_{\mu} \omega$ with $c_{\mu}=0.09$.
The length scale $l_{LES}$ is defined as
\begin{align}
            l_{\text{LES}} = C_{\text{DES}} \Delta , 
\end{align}
with the empirical constant $c_{DES}$ depending on the RANS background model. For the novel RSM-IDDES  $c_{DES}$ has been calibrated to $c_{DES}=0.65$.

\paragraph{WMLES Branch}
This branch of the IDDES method is active, if the inflow conditions contain resolved turbulence. The corresponding integral length scale is defined as:
\begin{equation}    l_{\text{WMLES}}=f_B(1+f_e)l_{\text{RANS}} + (1-f_B)l_{\text{LES}}
\end{equation}
The blending function $f_B$ as well as the elevating function $f_e$ represent empirical functions:
\begin{align}
\label{eqn:f_B}
f_B  &= \min{\{   2\exp(-9\alpha_g^2), 1.0 \}}, \ \ \   
\alpha_g = 0.25 -d_w/h_{\max} \\
f_e &= \max{  \{  (f_{e1} -1),0  \} }  f_{e2} \\
&\text{with}  \ \ \  f_{e1}\left( d_w / h_{max}\right) = 
\renewcommand\arraystretch{1.2}
\left\{
\begin{matrix}
2 \exp \left(-11.09\alpha_g^2\right)  \ \ \ &\text{if} \ \ \ \alpha_g \geq 0 \\
2 \exp \left(-9\alpha_g^2 \right) \ \ \ &\text{if} \ \ \ \alpha_g < 0
\end{matrix}
\right.
\renewcommand\arraystretch{1.0} 
\\
&\text{and} \ \ \    f_{e2}= 1.0 - \max{\{   f_t, f_l  \}} \ \  .
\end{align}
The function $f_{e2}$ influences the strength of the elevating function $f_e$ with the aid of the functions $f_t$ and $f_l$:
\begin{align}
    f_t&= \tanh{ \left[   {\left(c_t^2 r_{\text{dt}} \right)}^3  \right] }, \ \  f_l= \tanh{ \left[   {\left(c_l^2 r_{\text{dl}} \right)}^{10}  \right] } \\
\text{with} \ \ \  r_{\text{dt}} &= \frac{\nu_t}{ \kappa^2 d^2_w \cdot    {[ \sum_{i,j} {\left( \partial u_i/ \partial x_j \right)}^2 ]}^{1/2}  },  \ \ \ r_{\text{dl}} = \frac{\nu}{ \kappa^2 d^2_w \cdot    {[ \sum_{i,j} {\left( \partial u_i/ \partial x_j \right)}^2 ]}^{1/2}  }  \ \ .
\end{align}
Similar to $r_d$, the quantities $r_{dt}$ and $r_{dl}$ are markers of the turbulent boundary layer  and characterise the log layer ($r_{dt}=1$) and the laminar sublayer ($r_{dl}=1$), respectively.
The calibrations constants  $c_l$ and $c_t$ depend on the underlying RANS sub-grid model and amount to $c_l=5$ and $c_t=1.87$ for the novel RSM-IDDES.
\paragraph{Blending of the DDES and WMLES branches}
An automated switch between the DDES and WMLES branches depending on the inflow condition achieved by expressing the integral length scale as follows.
\begin{align}\label{eqn:lhyb_IDDES}
l_{hyb}  &= \tilde f_d \left( {1 + f_e } \right)l_{\mbox{\tiny RANS}}  + \left( {1 - \tilde f_d } \right)l_{\mbox{\tiny LES}}  \\
\label{eq:fdtilde}&\text{with} \ \ \    \tilde{f}_d =\max{ \{   (1-f_{dt}), f_B \} } \\
&\text{and} \ \ \     f_{dt}=1- \tanh{\left[ {\left(16r_{dt}\right)}^3    \right]}  \ \ \ .
\end{align}

\bibliographystyle{elsarticle-num}
\bibliography{elsarticle}

\begin{thebibliography}{10}
\expandafter\ifx\csname url\endcsname\relax
  \def\url#1{\texttt{#1}}\fi
\expandafter\ifx\csname urlprefix\endcsname\relax\def\urlprefix{URL }\fi
\expandafter\ifx\csname href\endcsname\relax
  \def\href#1#2{#2} \def\path#1{#1}\fi

\bibitem{Frohlich2008}
J.~Fr{\"{o}}hlich, D.~von Terzi,
  \href{http://dx.doi.org/10.1016/j.paerosci.2008.05.001}{{Hybrid LES/RANS
  Methods for the Simulation of Turbulent Flows}}, Progress in Aerospace
  Sciences 44~(5) (2008) 349--377.
\newline\urlprefix\url{http://dx.doi.org/10.1016/j.paerosci.2008.05.001}

\bibitem{spalart1997comments}
P.~R. Spalart, {Comments on the feasibility of LES for wings, and on a hybrid
  RANS/LES approach}, in: Proceedings of first AFOSR international conference
  on DNS/LES, Greyden Press, 1997.

\bibitem{Spalart2006}
P.~R. Spalart, S.~Deck, M.~L. Shur, K.~D. Squires, M.~K. Strelets, A.~Travin,
  {A New Version of Detached-Eddy Simulation, Resistant to Ambiguous Grid
  Densities}, Theoretical and Computational Fluid Dynamics 20~(3) (2006)
  181--195.
\newblock \href {https://doi.org/10.1007/s00162-006-0015-0}
  {\path{doi:10.1007/s00162-006-0015-0}}.

\bibitem{shur2008hybrid}
M.~L. Shur, P.~R. Spalart, M.~K. Strelets, A.~K. Travin, {A hybrid RANS-LES
  approach with delayed-DES and wall-modelled LES capabilities}, International
  journal of heat and fluid flow 29~(6) (2008) 1638--1649.

\bibitem{Masini2020a}
L.~Masini, S.~Timme, A.~J. Peace, {Scale-resolving simulations of a civil
  aircraft wing transonic shock-buffet experiment}, AIAA Journal 58~(10) (2020)
  4322--4338.
\newblock \href {https://doi.org/10.2514/1.J059219}
  {\path{doi:10.2514/1.J059219}}.

\bibitem{Probst2022}
A.~Probst, S.~Melber-Wilkending, {Hybrid RANS/LES of a generic high-lift
  aircraft configuration near maximum lift}, International Journal of Numerical
  Methods for Heat {\&} Fluid Flow 32~(4) (2022) 1204--1221.
\newblock \href {https://doi.org/10.1108/hff-08-2021-0525}
  {\path{doi:10.1108/hff-08-2021-0525}}.

\bibitem{Deck2012}
S.~Deck, {Recent improvements in the zonal detached eddy simulation (ZDES)
  formulation}, Theoretical and Computational Fluid Dynamics 26~(6) (2012)
  523--550.

\bibitem{Menter2018}
F.~Menter, {Stress-blended eddy simulation (SBES)—A new paradigm in hybrid
  RANS-LES modeling}, Notes on Numerical Fluid Mechanics and Multidisciplinary
  Design 137 (2018) 27--37.
\newblock \href {https://doi.org/10.1007/978-3-319-70031-1\_3}
  {\path{doi:10.1007/978-3-319-70031-1\_3}}.

\bibitem{Deck2020}
S.~Deck, N.~Renard, {Towards an enhanced protection of attached boundary layers
  in hybrid RANS/LES methods}, Journal of Computational Physics 400 (2020).
\newblock \href {https://doi.org/10.1016/j.jcp.2019.108970}
  {\path{doi:10.1016/j.jcp.2019.108970}}.

\bibitem{Nikitin2000}
N.~V. Nikitin, F.~Nicoud, B.~Wasistho, K.~D. Squires, P.~R. Spalart,
  \href{http://adsabs.harvard.edu/abs/2000PhFl...12.1629N}{{An Approach to Wall
  Modeling in Large-Eddy Simulations}}, Physics of Fluids 12~(7) (2000) 1629.
\newline\urlprefix\url{http://adsabs.harvard.edu/abs/2000PhFl...12.1629N}

\bibitem{Mockett2015}
C.~Mockett, M.~Fuchs, A.~Garbaruk, M.~Shur, P.~R. Spalart, M.~Strelets,
  F.~Thiele, A.~Travin, {Two non-zonal approaches to accelerate RANS to LES
  transition of free shear layers in DES}, in: Progress in Hybrid RANS-LES
  Modelling, Notes on Numerical Fluid Mechanics and Multidisciplinary Design,
  Vol. 130, 2015, pp. 187--201.

\bibitem{Kok2017}
J.~C. Kok, {A Stochastic Backscatter Model for Grey-Area Mitigation in Detached
  Eddy Simulations}, Flow, Turbulence and Combustion 99~(1) (2017) 119--150.
\newblock \href {https://doi.org/10.1007/s10494-017-9809-y}
  {\path{doi:10.1007/s10494-017-9809-y}}.

\bibitem{Probst2016}
A.~Probst, J.~L{\"{o}}we, S.~Reu{\ss}, T.~Knopp, R.~Kessler, {Scale-Resolving
  Simulations with a Low-Dissipation Low-Dispersion Second-Order Scheme for
  Unstructured Flow Solvers}, AIAA Journal 54~(10) (2016) 2972--2987.

\bibitem{Pont2017}
G.~Pont, P.~Brenner, P.~Cinnella, B.~Maugars, J.-C. Robinet,
  \href{http://dx.doi.org/10.1016/j.jcp.2017.08.036}{{Multiple-correction
  hybrid k-exact schemes for high-order compressible RANS-LES simulations on
  fully unstructured grids}}, Journal of Computational Physics 350 (2017)
  45--83.
\newblock \href {https://doi.org/10.1016/j.jcp.2017.08.036}
  {\path{doi:10.1016/j.jcp.2017.08.036}}.
\newline\urlprefix\url{http://dx.doi.org/10.1016/j.jcp.2017.08.036}

\bibitem{Spalart1992}
P.~R. Spalart, S.~R. Allmaras, {A One-Equation Turbulence Model for Aerodynamic
  Flows}, in: AIAA Paper 92-0439, 1992, p. 439.

\bibitem{Menter1994}
F.~R. Menter, {Two-Equation Eddy-Viscosity Turbulence Models for Engineering
  Applications}, AIAA journal 32~(8) (1994) 1598--1605.

\bibitem{Eisfeld2016}
B.~Eisfeld, C.~Rumsey, V.~Togiti, {Verification and validation of a
  second-moment-closure model}, AIAA Journal 54~(5) (2016) 1524--1541.
\newblock \href {https://doi.org/10.2514/1.J054718}
  {\path{doi:10.2514/1.J054718}}.

\bibitem{Jakirlic2016}
S.~Jakirli{\'{c}}, R.~Maduta, {“Steady” RANS Modeling for Improved
  Prediction of Wall-Bounded Separation}, AIAA Journal 54~(5) (2016)
  1803--1809.
\newblock \href {https://doi.org/10.2514/1.J054399}
  {\path{doi:10.2514/1.J054399}}.

\bibitem{SPORSCHILL2022108955}
G.~Sporschill, F.~Billard, M.~Mallet, R.~Manceau, H.~Bézard, {Assessment of
  Reynolds-stress models for aeronautical applications}, International Journal
  of Heat and Fluid Flow 96 (2022) 108955.
\newblock \href
  {https://doi.org/https://doi.org/10.1016/j.ijheatfluidflow.2022.108955}
  {\path{doi:https://doi.org/10.1016/j.ijheatfluidflow.2022.108955}}.

\bibitem{francois2015forced}
D.~G. Francois, R.~Radespiel, A.~Probst, {Forced synthetic turbulence approach
  to stimulate resolved turbulence generation in embedded LES}, Notes on
  Numerical Fluid Mechanics and Multidisciplinary Design 130 (2015) 81--92.
\newblock \href {https://doi.org/10.1007/978-3-319-15141-0\_6}
  {\path{doi:10.1007/978-3-319-15141-0\_6}}.

\bibitem{Ehrle2020}
M.~Ehrle, A.~Waldmann, T.~Lutz, E.~Kr{\"{a}}mer,
  \href{https://doi.org/10.1007/s13272-020-00466-7}{{Simulation of transonic
  buffet with an automated zonal DES approach}}, CEAS Aeronautical Journal
  11~(4) (2020) 1025--1036.
\newblock \href {https://doi.org/10.1007/s13272-020-00466-7}
  {\path{doi:10.1007/s13272-020-00466-7}}.
\newline\urlprefix\url{https://doi.org/10.1007/s13272-020-00466-7}

\bibitem{Probst2011}
A.~Probst, R.~Radespiel, T.~Knopp, {Detached-Eddy Simulation of Aerodynamic
  Flows Using a Reynolds-Stress Background Model and Algebraic RANS / LES
  Sensors}, in: AIAA Paper 2011-3206, 2011, p. 3206.

\bibitem{Maduta2012}
R.~Maduta, S.~Jakirlić, {An Eddy-Resolving Reynolds Stress Transport Model for
  Unsteady Flow Computations}, Notes on Numerical Fluid Mechanics and
  Multidisciplinary Design 117 (2012) 77--89.
\newblock \href {https://doi.org/10.1007/978-3-642-31818-4-6}
  {\path{doi:10.1007/978-3-642-31818-4-6}}.

\bibitem{Zhuchkov2016a}
R.~N. Zhuchkov, A.~A. Utkina, {Combining the SSG/LRR-$\omega$ differential
  reynolds stress model with the detached eddy and laminar-turbulent transition
  models}, Fluid Dynamics 51~(6) (2016) 733--744.
\newblock \href {https://doi.org/10.1134/S001546281606003X}
  {\path{doi:10.1134/S001546281606003X}}.

\bibitem{chaouat2005new}
B.~Chaouat, R.~Schiestel, {A new partially integrated transport model for
  subgrid-scale stresses and dissipation rate for turbulent developing flows},
  Physics of Fluids 17~(6) (2005) 065106.

\bibitem{Probst2015a}
A.~Probst, S.~Reu{\ss}, {Scale-Resolving Simulations of Wall-Bounded Flows with
  an Unstructured Compressible Flow Solver}, in: Progress in Hybrid RANS-LES
  Modelling. Notes on Numerical Fluid Mechanics and Multidisciplinary Design,
  Springer International Publishing, Vol. 130, 2015, pp. 481--491.

\bibitem{Wang2021}
G.~Wang, Q.~Li, Y.~Liu, \href{https://doi.org/10.1016/j.ast.2021.107207}{{IDDES
  method based on differential Reynolds-stress model and its application in
  bluff body turbulent flows}}, Aerospace Science and Technology 119 (2021)
  107207.
\newblock \href {https://doi.org/10.1016/j.ast.2021.107207}
  {\path{doi:10.1016/j.ast.2021.107207}}.
\newline\urlprefix\url{https://doi.org/10.1016/j.ast.2021.107207}

\bibitem{Carlsson2023}
M.~Carlsson, L.~Davidson, S.~H. Peng, S.~Arvidson,
  \href{https://doi.org/10.1016/j.compfluid.2022.105741}{{Investigation of
  low-dissipation low-dispersion schemes for incompressible and compressible
  flows in scale-resolving simulations}}, Computers and Fluids 251~(November
  2022) (2023) 105741.
\newblock \href {https://doi.org/10.1016/j.compfluid.2022.105741}
  {\path{doi:10.1016/j.compfluid.2022.105741}}.
\newline\urlprefix\url{https://doi.org/10.1016/j.compfluid.2022.105741}

\bibitem{Schwamborn2006}
D.~Schwamborn, T.~Gerhold, R.~Heinrich, {The DLR TAU-Code: Recent Applications
  in Research and Industry}, in: M.~Braza, A.~Bottaro, M.~Thompson (Eds.),
  ECCOMAS CFD, P. Wesseling, E. O{\~{n}}ate, J. P{\'{e}}riaux (Eds), TU Delft,
  The Netherlands, 2006.

\bibitem{eisfeld2022reynolds}
B.~Eisfeld, C.~L. Rumsey, V.~Togiti, S.~Braun, A.~St{\"u}rmer, {Reynolds-Stress
  Model Computations of NASA Juncture Flow Experiment}, AIAA Journal 60~(3)
  (2022) 1643--1662.

\bibitem{braun2019implementation}
S.~Braun, {Implementation of a ln ($\omega$)-based SSG/LRR Reynolds Stress
  Model into the DLR-TAU Code}, DLR report No. IB-AS-BS-2019-37 (2019).

\bibitem{Loewe2016}
J.~L{\"{o}}we, A.~Probst, T.~Knopp, R.~Kessler, {Low-Dissipation Low-Dispersion
  Second-Order Scheme for Unstructured Finite-Volume Flow Solvers}, AIAA
  Journal 54~(10) (2016) 2961--2971.

\bibitem{kok2009high}
J.~Kok, {A high-order low-dispersion symmetry-preserving finite-volume method
  for compressible flow on curvilinear grids}, Journal of Computational Physics
  228~(18) (2009) 6811--6832.

\bibitem{Adamian2011a}
D.~Adamian, A.~Travin, {An Efficient Generator of Synthetic Turbulence at
  RANS-LES Interface in Embedded LES of Wall-Bounded and Free Shear Flows}, in:
  A.~Kuzmin (Ed.), Computational Fluid Dynamics 2010, Springer Berlin
  Heidelberg, 2011, pp. 739--744.
\newblock \href {https://doi.org/10.1007/978-3-642-17884-9}
  {\path{doi:10.1007/978-3-642-17884-9}}.

\bibitem{Shur2014}
M.~Shur, P.~Spalart, M.~Strelets, A.~Travin, {Synthetic Turbulence Generators
  for RANS-LES Interfaces in Zonal Simulations of Aerodynamic and Aeroacoustic
  Problems}, Flow Turbulence and Combustion 93 (2014) 63--92.
\newblock \href {https://doi.org/10.1007/s10494-014-9534-8}
  {\path{doi:10.1007/s10494-014-9534-8}}.

\bibitem{Probst2020}
A.~Probst, P.~Str{\"{o}}er, {Comparative Assessment of Synthetic Turbulence
  Methods in an Unstructured Compressible Flow Solver}, Notes on Numerical
  Fluid Mechanics and Multidisciplinary Design 143 (2020) 193--202.
\newblock \href {https://doi.org/10.1007/978-3-030-27607-2\_15}
  {\path{doi:10.1007/978-3-030-27607-2\_15}}.

\bibitem{comte1971simple}
G.~Comte-Bellot, S.~Corrsin, {Simple Eulerian time correlation of full-and
  narrow-band velocity signals in grid-generated,‘isotropic’turbulence},
  Journal of fluid mechanics 48~(2) (1971) 273--337.

\bibitem{Kraichnan1970}
R.~H. Kraichnan, {Diffusion by a Random Velocity Field}, The Physics of Fluids
  13~(1) (1970) 22--31.

\bibitem{probst2016scale}
A.~Probst, J.~L{\"o}we, S.~Reu{\ss}, T.~Knopp, R.~Kessler, {Scale-resolving
  simulations with a low-dissipation low-dispersion second-order scheme for
  unstructured flow solvers}, AIAA Journal (2016) 2972--2987.

\bibitem{moser1999direct}
R.~D. Moser, J.~Kim, N.~N. Mansour, {Direct numerical simulation of turbulent
  channel flow up to Re $\tau$= 590}, Physics of fluids 11~(4) (1999) 943--945.

\bibitem{lozano2014effect}
A.~Lozano-Dur{\'a}n, J.~Jim{\'e}nez, {Effect of the computational domain on
  direct simulations of turbulent channels up to Re $\tau$= 4200}, Physics of
  Fluids 26~(1) (2014) 011702.

\bibitem{gritskevich2012development}
M.~S. Gritskevich, A.~V. Garbaruk, J.~Sch{\"u}tze, F.~R. Menter, et~al.,
  {Development of DDES and IDDES formulations for the k-$\omega$ shear stress
  transport model}, Flow Turbulence and Combustion 88~(3) (2012) 431.

\bibitem{Deck2018}
S.~Deck, P.-e. Weiss, N.~Renard,
  \href{https://doi.org/10.1016/j.jcp.2018.02.028}{{A rapid and low noise
  switch from RANS to WMLES on curvilinear grids with compressible flow
  solvers}}, Journal of Computational Physics 363 (2018) 231--255.
\newblock \href {https://doi.org/10.1016/j.jcp.2018.02.028}
  {\path{doi:10.1016/j.jcp.2018.02.028}}.
\newline\urlprefix\url{https://doi.org/10.1016/j.jcp.2018.02.028}

\bibitem{nagib2007approach}
H.~M. Nagib, K.~A. Chauhan, P.~A. Monkewitz, {Approach to an asymptotic state
  for zero pressure gradient turbulent boundary layers}, Philosophical
  Transactions of the Royal Society A: Mathematical, Physical and Engineering
  Sciences 365~(1852) (2007) 755--770.

\bibitem{Shur2017}
M.~Shur, M.~Strelets, A.~Travin, A.~Probst, S.~Probst, D.~Schwamborn, S.~Deck,
  A.~Skillen, J.~Holgate, A.~Revell, {Improved Embedded Approaches}, in:
  Go4Hybrid: Grey Area Mitigation for Hybrid RANS-LES Methods, Notes on
  Numerical Fluid Mechanics and Multidisciplinary Design 134, Springer
  International Publishing, 2017, pp. 51--87.

\end{thebibliography}

\end{document}